\documentclass[10pt,prx,aps,twocolumn,showpacs,floatfix]{revtex4}
\usepackage{graphicx}
\usepackage{amsmath}
\usepackage{bm} 
\usepackage{color}
\usepackage{lipsum}
\usepackage{braket} 
\usepackage{booktabs}

\begin{document}
\title{Non-equilibrium quantum mechanics: A `hot quantum soup' of paramagnons
}
\author{H. D. Scammell and O. P. Sushkov }
\affiliation{School of Physics, The University of New South Wales,
  Sydney, NSW 2052, Australia}
\date{\today}
\begin{abstract}
Motivated by recent measurements of the lifetime (decay width) of paramagnons in quantum 
antiferromagnet TlCuCl$_3$, we investigate paramagnon
decay in a heat bath and formulate an appropriate quantum theory. Our formulation can be split into two regimes: (i) a non-perturbative, `hot quantum soup' regime where paramagnon width is comparable to its energy; (ii) usual perturbative regime where paramagnon width is significantly lower than its energy.
Close to the Neel temperature the paramagnon width becomes comparable
to its energy and falls into the hot quantum soup regime. To describe this regime
we develop a new finite frequency, finite temperature technique
for a nonlinear quantum field theory; the `golden rule of quantum kinetics'.
The formulation is generic and applicable to any three dimensional
quantum antiferromagnet in the vicinity of a quantum critical point. Specifically we apply our results to TlCuCl$_3$ and find agreement with experimental 
data. 
Additionally, we show that logarithmic running of the coupling constant in
the upper critical dimension changes the commonly accepted picture
of the quantum disordered and quantum critical regimes.

\end{abstract}

\maketitle

\section{Introduction}
Understanding the interplay between thermal and quantum fluctuations in quantum systems is an exciting challenge to theory. In particular, understanding how to appropriately treat (quasi-) particles in a hot and dense medium is of fundamental importance to many areas of physics ranging from condensed matter, to plasma, nuclear, and particle physics. In this work we concentrate on lifetimes of quasiparticles, or, more generally, on line-shapes of spectral functions. The lifetime and the spectral function are essentially non-equilibrium properties in spite of the fact that the entire many-body system that we consider is in thermal equilibrium. A perturbative treatment of quasiparticles in a hot dense medium becomes plagued by infrared divergences that occur due to the medium. In this paper we develop and present a relatively simple technique that i) regulates the infrared behavior via a resummation of medium effects i.e. the self-consistent inclusion of line-shapes, and ii) allows one to handle the calculation of non-equilibrium responses at finite temperature.

\begin{figure}[h!]
\includegraphics[width=0.35\textwidth,clip]{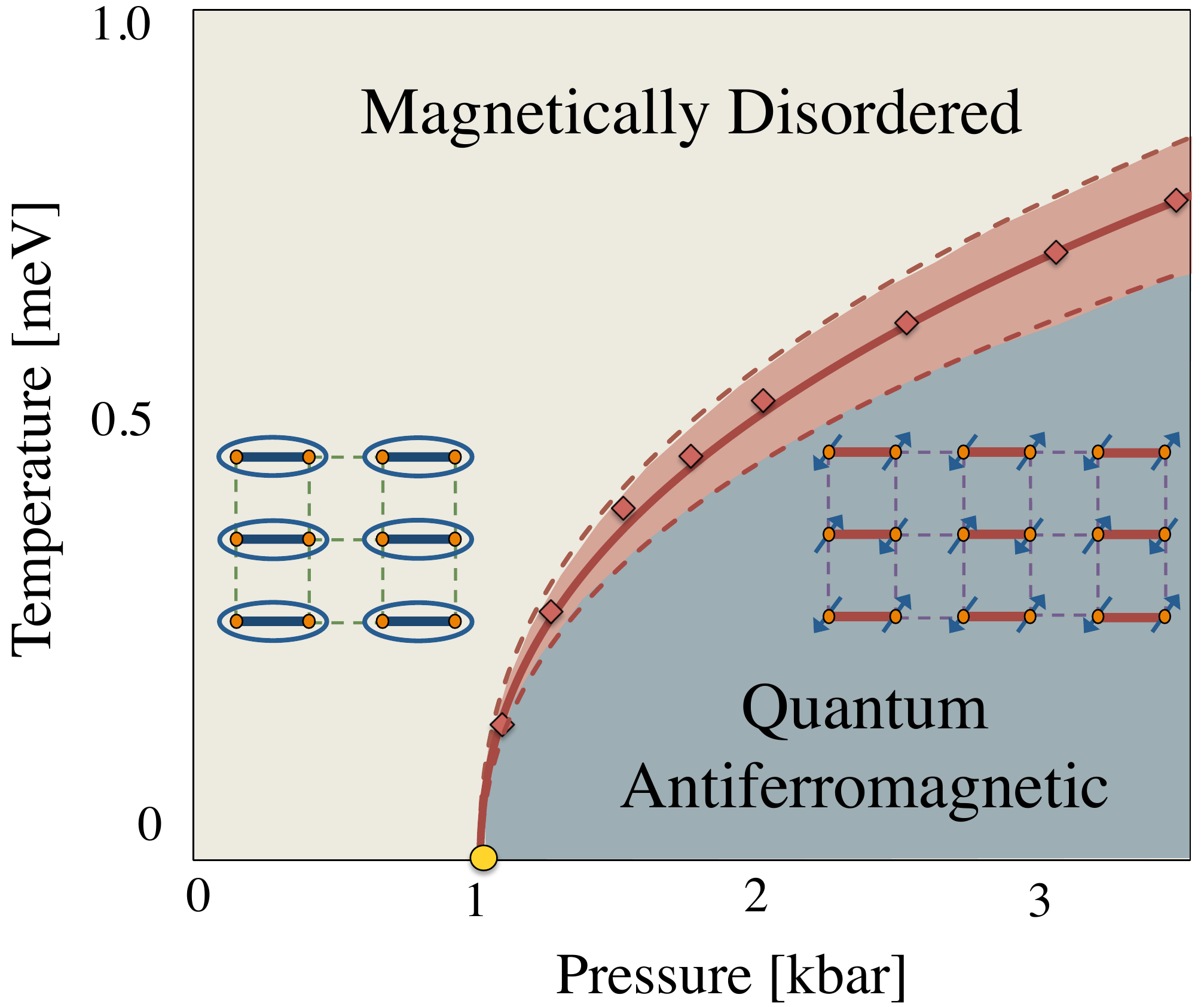}
\caption{ The pressure-temperature phase diagram of  TlCuCl$_3$.
The QCP is at $p=p_c=1.01$kbar.
The N\'eel temperature curve separates
magnetically ordered and magnetically disordered phases.
The light red band around the N\'eel curve indicates the region of dimensional 
crossover.
}
\label{PD}
\end{figure}

The problem we investigate was stimulated by the observation of 
paramagnons
in the magnetically disordered phase of the three dimensional (3D),  
dimerized quantum antiferromagnet TlCuCl$_3$~\cite{Merchant}. 
The pressure-temperature phase diagram of the compound is shown in 
Figure (\ref{PD}).
The quantum phase transition at the quantum critical point (QCP)
$p=p_c=1.01$kbar is driven by external hydrostatic
pressure. The red line in Fig.\ref{PD} shows the N\'eel temperature versus  
pressure~\cite{Ruegg}. At $p > p_c$ and temperatures below the N\'eel curve, 
the compound possesses long range
antiferromagnetic order. Going above the N\'eel curve at $p>p_c$, the system becomes magnetically disordered, while
at $p < p_c$ the system is disordered even at zero temperature.
Magnetic excitations at zero temperature and at $p < p_c$ are usually called 
triplons,
while magnetic excitations at $p > p_c$ and $T > T_N$ are usually called 
paramagnons.
It is clear from Fig.~\ref{PD} that there is no qualitative difference 
between triplons and 
paramagnons and so throughout this work we will exclusively use the 
term paramagnon, {\textit i.e}  a triplon is a paramagnon.

It was observed~\cite{Merchant,Ruegg}  that at temperatures just above 
the N\'eel temperature $T_N$,  the paramagnons are relatively broad  
$\Gamma/\omega \gtrsim 1$, here $\Gamma$ is the width and $\omega$ is 
the energy of the paramagnon. At increasing temperatures, the paramagnons 
become narrow, $\Gamma/\omega \ll 1$. 
This unexpected behaviour is an indication of a nontrivial 
interplay between quantum and thermal fluctuations~\cite{sushkov}.

While  TlCuCl$_3$ is a spin dimerized compound,
the phase diagram in  Figure (\ref{PD}) is essentially the generic phase
diagram  of a 3D isotropic quantum antiferromagnet~\cite{Sachdev11}, dimerized or not.
The widths of magnons in the magnetically ordered phase of quantum
magnets have received both theoretical
and experimental 
attention~\cite{Harris,Tyc,Kopietz,Bayrakci,ZhitomirskyReview}.
On the other hand we are not aware of any previous theoretical studies of decay widths of paramagnons in the disordered phase of 3D quantum 
antiferromagnets at finite temperatures.

In the magnetically ordered phase  at low temperatures, $T < T_N$, there exists two
types of magnetic excitations. First there are Goldstone excitations called magnons.
Magnons are generally long lived quasiparticles which weakly interact with each other~\cite{Harris,Tyc,Kopietz,Bayrakci}. 
This holds especially true for 
higher dimensional, non-frustrated systems, or systems without spontaneous 
decay \cite{ZhitomirskyReview}.
The long lifetime of magnons, $\Gamma/\omega \ll 1$,
 is due to Adler's theorem which claims that the magnon-magnon
interaction must vanish in the long wave-length limit. 
Adler's theorem is a general dynamic
property unrelated to the magnitude of the effective coupling constant.
Also within the magnetically ordered phase, along with the Goldstone magnons, there exist longitudinal (Higgs) magnetic excitations.
The width of Higgs excitations depends on the magnitude of the effective coupling 
constant, and is not governed by
Adler's theorem. It can be large, $\Gamma/\omega \gtrsim 1$, like in the Heisenberg model
on a simple square or cubic lattice, or it can be small,  $\Gamma/\omega \ll 1$, like in
TlCuCl$_3$~\cite{Kulik} and some other dimerzied spin systems.

In the present work we develop, and subsequently apply, a technique to 
calculate widths of paramagnons
in the magnetically disordered phase of a 3D quantum system
in the vicinity of a QCP.
While specifically we discuss an O(3) field theory (and apply to the real 
compound TlCuCl$_3$), the developed techniques are generic and are applicable
to all systems of this kind; symmetric phases described by O(N)-field theories. 
For example, they are applicable to the electroweak phase transition in 
cosmology; to the wide class of spin dimerized magnetic models~\cite{Qin}; and to O(2) superfluids or superconductors in the vicinity of their QCP's. 

The paper is organized as follows; in Section \ref{GenCon} we introduce the 
necessary mathematical and physical techniques. Section \ref{Intuitive} provides 
an intuitive picture of the decay and scattering processes, with particular 
focus on the influence of a heat bath. 
Section \ref{QDQC} addresses quantum disordered and quantum critical
regimes. We show that they are somewhat different from the commonly accepted
picture.
Section \ref{Conjecture} discusses the inconsistency of the usual perturbative Fermi golden rule, and introduces our proposed `golden rule of quantum kinetics', which simultaneously incorporates decay and heat bath scattering processes, as well as providing a self-consistent, nonequilibrium technique to calculate widths. A general mathematical analysis of the golden rule of quantum kinetics, without reference to any particular system, is given in Section \ref{GeneralResults}. Finally in Section \ref{Comparison} we apply our technique to the specific compound TlCuCl$_3$, and compare our results with 
inelastic neutron scattering  experimental data. 

\section{General considerations}\label{GenCon}
In the vicinity of the quantum critical point, quantum antiferromagnets are described by the Landau-Ginzburg-like 
effective field theory \cite{Sachdev11, Affleck}
\begin{align}
\label{DisorderedLagrangian}
{\cal L}&=\frac{1}{2}\partial_{\mu}\vec\varphi\partial^\mu\vec\varphi - \frac{1}{2}m_0^2\vec\varphi^2-\frac{1}{4}\alpha_0\vec\varphi^4,
\end{align}
where $\vec\varphi=(\varphi_1,\varphi_2,\varphi_3)$ is a three component real 
vector field  describing  the spin $S=1$ magnetic excitations. 
The index, $\mu=0,x,y,z$, enumerates time and three-space coordinates, and the paramagnon speed is set equal to unity, $c=1$.
The bare coupling constant is $\alpha_0$,
and the bare effective mass squared $m_0^2$ changes sign at the QCP,
$m_0^2 = \gamma^2(g_c-g)$, where $g$ is some external parameter and
$\gamma$ is a coefficient.
For example in TlCuCl$_3$ the transition is driven by external pressure,
$m_0^2=\gamma^2(p_c-p)$.
Below we use the rescaled coupling constant,
\begin{equation}
\label{b}
\beta=\frac{\alpha}{8\pi}\ ,
\end{equation}
it is a more natural combination for perturbation theory.
To apply perturbation theory and the renormalization group (RG)
we assume that $\beta \ll 1$.
This is always true in a sufficiently close vicinity of the QCP.
Quantum and thermal fluctuations lead to running of both the coupling 
constant and the effective mass; they become energy,
momentum, and temperature dependent, $\beta_0 \to \beta_q$, $m_0^2 \to m_q^2$.
Equations for these quantities, derived in
Ref.~\cite{Scammell}, are valid everywhere in the phase diagram Fig.~\ref{PD}.
In the present work we calculate the width and spectral function
of paramagnons within the magnetically disordered region of the phase diagram 
Fig.~\ref{PD}.

As a mathematical object we use the retarded Green's function of the paramagnon,
which is an analytic continuation of the Matsubara Green's function
from the upper imaginary energy half-axis to the real energy axis. 
To have a coherent presentation
we remind here basic properties of the retarded Green's function 
$G^R(\omega, q)$, see e.g. Ref.~\cite{Liftshitz}.
For the case of a noninteracting field, $\beta=0$, 
the Lagrangian (\ref{DisorderedLagrangian}) becomes
\begin{eqnarray} 
\label{L}
{\cal L}=\frac{1}{2}\partial_{\mu}\vec\varphi\partial^\mu\vec\varphi -\frac{1}{2}m_0^2{\vec \varphi}^2\ ,
\end{eqnarray}
and the exact Green's function is immediately deduced 
\begin{eqnarray}
\label{gr}
G^R(\omega,{\bf q})&=&\frac{1}{2\omega_q}\left\{\frac{1}{\omega-\omega_q+i0}
-\frac{1}{\omega+\omega_q+i0}\right\}\nonumber\\
\omega_q&=&\sqrt{q^2+m_0^2} \ .
\end{eqnarray}
This is true for both zero and nonzero temperatures, as soon as there is no 
interaction.
From (\ref{gr}) we see symmetry properties of $G^R$, the real part of $G^R$ 
is an even function of $\omega$ while the imaginary part of $G^R$ is odd.
These are general properties valid also in the case of non-zero interaction.

The general spectral representation of $G^R$ follows,
see Ref.~\cite{Liftshitz},
\begin{eqnarray}
\label{sg}
iG^R(x,0)&=&
\frac{1}{3}\sum_{nm}\frac{e^{-E_n/T}}{Z}
e^{-i\omega_{mn}t+i{\bm k}_{mn}\cdot{\bm r}}\\
&\times& \left\{1-e^{-\omega_{mn}/T}\right\}
|\langle m|\varphi_{i}(0)|n\rangle|^2 \ .\nonumber
\end{eqnarray}
Here $|n\rangle$ and $|m\rangle$ are exact stationary quantum states 
of the system, $E_n$ and ${\bm k}_n$ are
the energy and the momentum of the state,
$\omega_{mn}=E_m-E_n$, ${\bm k}_{mn}={\bm k}_m-{\bm k}_n$, while
 $Z$ is the partition function. 

Now consider the interaction of some external source $J_{i}$, 
with the paramagnon field $\varphi_{i}$ (for instance  $J_{i}$ can be the magnetic field of a neutron scattered from the system), 
\begin{equation}
\label{hint}
{\cal L}_{int}=J_{i}\varphi_{i} \ .
\end{equation}
Assuming that this interaction is very weak the 
probability $W$ of the system
 excitation per unit time, due to interaction with 
the external source (\ref{hint}), is given by the Fermi golden rule. 
\begin{eqnarray}
\label{fgr}
W\propto S_{\bm q}(\omega)&=&\frac{1}{3}\sum_{nm}\frac{e^{-E_n/T}}{Z}|\langle m|\varphi_{\alpha}(0)|n\rangle|^2\\
&\times&\delta(\omega -\omega_{mn})(2\pi)^3\delta({\bm q}-{\bm k}_{mn}) \ . 
\nonumber
\end{eqnarray}
Here $\omega$ is the energy transfer and ${\bm q}$ is the momentum transfer 
to the system.
So a scattering experiment allows one to measure the
structure factor $S_{\bm q}(\omega)$ defined by Eq.(\ref{fgr}).
Comparison of Eqs. (\ref{sg}) and (\ref{fgr}) results in the following,
important, exact relation
\begin{equation}
\label{GS}
-\frac{1}{\pi}Im \ G^R(\omega,{\bm q})=(1-e^{-\omega/T})S_ {\bm q}(\omega) \ .
\end{equation}
Note that Eqs. (\ref{sg}), (\ref{fgr}), and (\ref{GS}) are exact, they are valid
at arbitrary interaction and arbitrary temperature.  
Another exact theorem immediately follows from these equations; the imaginary
part of $G^R(\omega,{\bm q})$ is an odd function of $\omega$ as
already pointed out above.

Account of interaction $\alpha_0{\vec{\varphi}}^4/4$ in 
Eq.(\ref{DisorderedLagrangian})
leads to a paramagnon self-energy $\Sigma_q(\omega)$.
Of course the self-energy depends on temperature, however, for ease of notation
we do not write temperature as an explicit argument. 
The real part of the self-energy has been calculated earlier using the
single loop  renormalization group (RG)~\cite{Scammell}.
Account of the real part leads to the replacement $m_0^2\to m^2_q$
in  Eq.(\ref{gr}), where $m_q\equiv \Delta$ is the renormalized mass, such that the dispersion is given by
\begin{equation}
\label{oq}
\omega_q=\sqrt{q^2+\Delta^2} \ .
\end{equation}
\begin{figure}[h!]
\includegraphics[width=0.47\textwidth,clip]{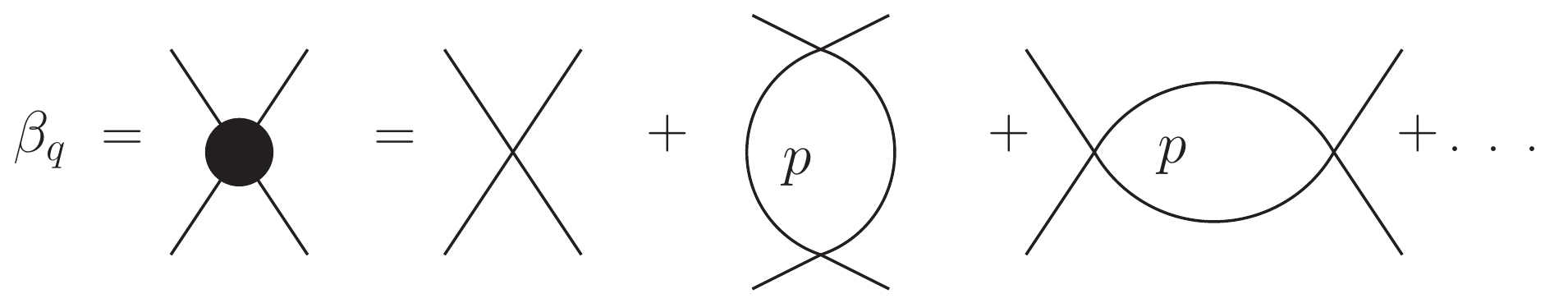}
\caption{ Diagrammatic subseries: Coupling constant.
}
\label{Fig22}
\end{figure}
\begin{figure}[h!]
\includegraphics[width=0.49\textwidth,clip]{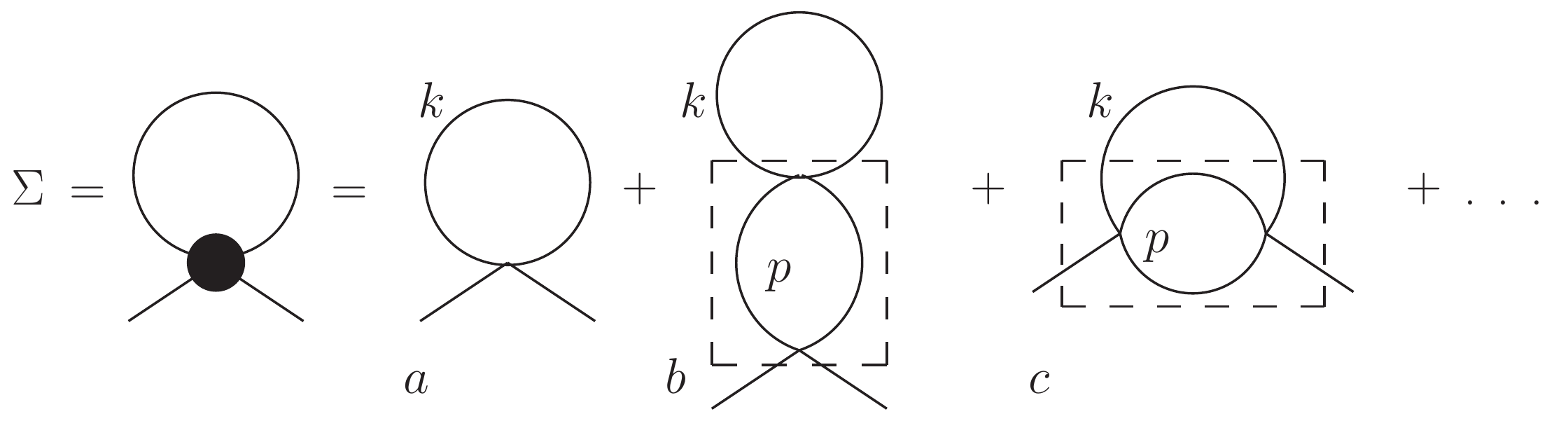}
\caption{ Diagrammatic subseries: Self-Energy.
}
\label{Fig33}
\end{figure}
\noindent Generally $\Delta$ depends momentum and temperature. 
 Below we take $\omega_q$ as given by Eq.(\ref{oq}).
{
It is important to understand the structure of diagrams included in the self energy.
The diagrams contributing to the running coupling constant $\beta_q$ are shown schematically in Fig.\ref{Fig22}.
The momentum in the loop runs in the limits $\Lambda_0 > p > q$, where q is the external momentum
and $\Lambda_0$ is the ultraviolet cutoff.
The self energy is given by diagrams shown schematically in Fig.\ref{Fig33}.
All diagrams are quadratically, ultraviolet divergent. Quadratic divergences have no physical meaning and are removed
during the renormalization. 
After removal of the quadratic divergence the typical momentum in the ``external'' loop is 
$k\sim \Delta, T$ while the typical
momentum in the ``internal'' loop is  $\Lambda_0 > p > \Delta,T$. 
The internal loops of the double loop diagrams are inside dashed boxes in Fig.\ref{Fig33}b,c. The series of internal loops can be identified as the series of the running coupling constant, as shown in Fig.\ref{Fig22}. 
The point to note is that most important logarithmically divergent  part of the
``sunset'' diagram (Fig.\ref{F4}, considered in the next section)
 is fully included in our RG calculation of $\Delta$~\cite{Scammell}. For example, Fig.\ref{Fig33}c is a part of the ``sunset" diagram.
In the diagrammatic series Fig.\ref{Fig33} we consider only the real part of the ``sunset'' diagram. A central point of this work is the consideration the imaginary part of the ``sunset" diagram. However, to extract the most important physics relating to the imaginary part, we will need to consider a different, infinite subseries. See Fig.\ref{F5}. The following sections are dedicated to this point.
}

The imaginary part of the self-energy describes broadening
\begin{eqnarray}
\label{br}
&&\Gamma_q(\omega)=-\frac{Im\Sigma_q(\omega)}{\omega}\\
&&G^R(\omega,{\bf q})=
\frac{1}{\omega^2-\omega_q^2-\Sigma_q(\omega)} \to
\frac{1}{\omega^2-\omega_q^2+i\omega\Gamma_q(\omega)} \ .\nonumber
\end{eqnarray}
There are two points to note, (i) generally $\Gamma_q$ depends on
$\omega$ and hence the line shape can be significantly different
from that of a simple Lorentzian; (ii) $\Gamma_q(\omega)$ is an even function of
$\omega$ since $Im\Sigma_q(\omega)$ is an odd function.
The structure factor corresponding to (\ref{br}) immediately
follows from Eq.(\ref{GS}),
\begin{eqnarray}
\label{spectral}
S_ {\bm q}(\omega)=
\frac{1}{\pi(1-e^{-\frac{\omega}{T}})}\left\{\frac{\omega\Gamma_q}
{[\omega^2-\omega_q^2]^2+\omega^2\Gamma_q^2}\right\}.
\end{eqnarray}

\section{Intuitive analysis and perturbation theory }\label{Intuitive}
Let $\Phi$ be a paramagnon for which we are determining the decay rate;
the ``probe paramagnon''.
The probe paramagnon can spontaneously decay into 3 paramagnons as
shown 
\begin{figure}[h!]
\includegraphics[width=0.3\textwidth,clip]{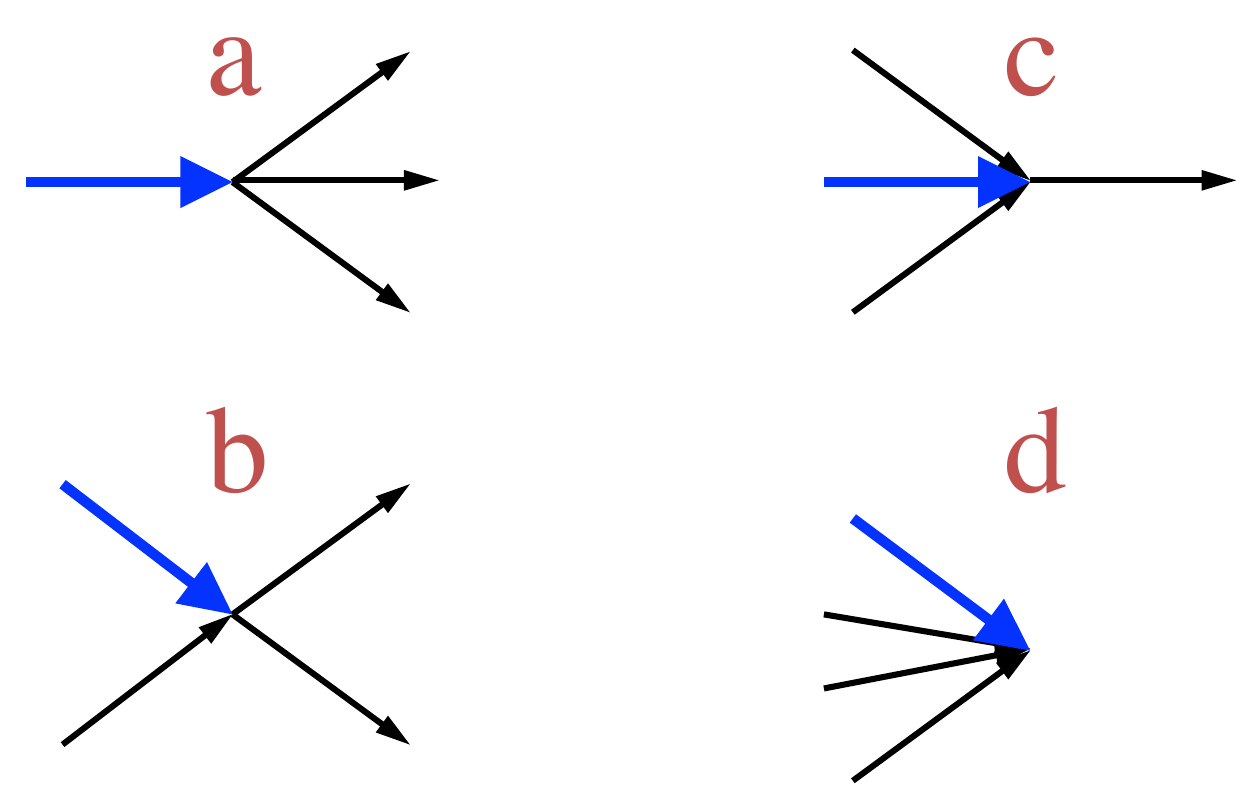}
\caption{ ``Decay'' diagrams for a paramagnon
 The thick blue line represents 
the probe
paramagnon and thin black lines represent the heat bath paramagnons.
}
\label{F2}
\end{figure}
in Fig.\ref{F2}a.
In the presence of a heat bath, the probe paramagnon can also scatter 
from a bath paramagnon - this is the Raman process 
shown in  Fig.\ref{F2}b.
The fusion process with two or even three heat bath paramagnons
is also possible, Figs. \ref{F2}c and \ref{F2}d.
It is worth noting that processes Fig.\ref{F2}a,c,d
are kinematically forbidden for on-mass-shell paramagnons with
dispersion (\ref{oq}) \cite{ordered phase}. However, one must include 
the processes in the analysis
because close to the N\'eel temperature paramagnons are broad and the
mass-shell notion is not defined. 

\begin{figure}[h!]
\includegraphics[width=0.3\textwidth,clip]{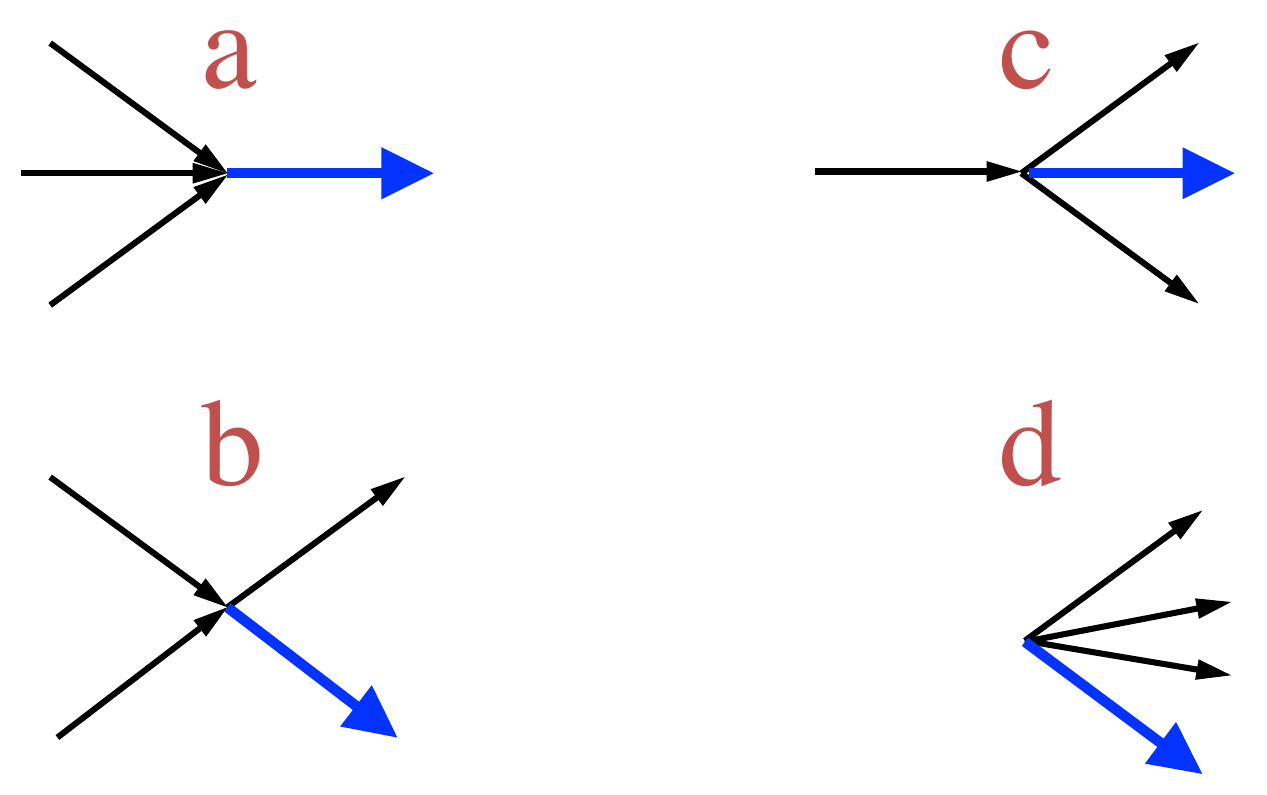}
\caption{ Diagrams corresponding to pumping (inverse processes)
to the paramagnon state. 
The thick blue line represents 
the probe paramagnon and thin black lines represent the heat bath paramagnons.
}
\label{F3}
\end{figure}

Along with each of the above four decay processes, there also exists 
their inverse process - ``pumping" from the paramagnon bath shown in 
Fig.\ref{F3}.
It is intuitively clear that 
\begin{equation}
\label{ggg}
\Gamma_q(\omega)=\Gamma_q^{(d)}(\omega)-\Gamma^{(i)}_q(\omega) \ ,
\end{equation}
where $\Gamma_q$ is the total width in Eq.(\ref{br}), $\Gamma^{(d)}_q$ is the
decay width associated with processes in Fig.\ref{F2}
and $\Gamma^{(i)}_q$ is the inverse width associated with processes in 
Fig.\ref{F3}. For a formal derivation of (\ref{ggg}) see Ref.~\cite{Weldon}.
Due to the detailed balance there is a simple relation between the decay
and the inverse widths~\cite{Weldon, Holstein}. 
\begin{eqnarray}
&&\Gamma^{(i)}_q(\omega)=e^{-\omega/T}\Gamma^{(d)}_q(\omega)\nonumber\\
\label{di1}
&&\Gamma_q(\omega)=(1-e^{-\omega/T})\Gamma^{(d)}_q(\omega) \ .
\end{eqnarray}
It is interesting to note that while relation (\ref{ggg}) is valid for bosons,
for fermions $\Gamma=\Gamma^{(d)}+\Gamma^{(i)}$, see Ref.~\cite{Weldon}.

Now we look at simple perturbation theory which is equivalent to
the Fermi golden rule. 
Direct application of Fermi Golden rule to diagrams in Fig.~\ref{F2}
gives the following decay width
\begin{eqnarray}
\label{gd}
\Gamma^{(d)}_{q}(\omega)&=&\frac{16(2\pi)^6\mathcal{S}\beta_0^2}{2\omega}
\int\frac{d^3k_1}{2\omega_1(2\pi)^3}\frac{d^3k_2}{2\omega_2(2\pi)^3}
\frac{d^3k_3}{2\omega_3(2\pi)^3}\nonumber\\
&&\hspace{-15pt}\times[(1+n_{1})(1+n_{2})(1+n_{3}) \ \delta^{(4)}(q-k_1-k_2-k_3)\nonumber\\
&&\hspace{-15pt}+3n_{1}(1+n_{2})(1+n_{3})\ \delta^{(4)}(q+k_1-k_2-k_3)\nonumber\\
&&\hspace{-15pt}+3n_{1}n_{2}(1+n_{3})\ \delta^{(4)}(q+k_1+k_2-k_3)\nonumber\\
&&\hspace{-15pt}+n_{1}n_{2}n_{3}\ \delta^{(4)}(q+k_1+k_2+k_3)]
\end{eqnarray}
Here 
\begin{equation}
\label{BE}
n_k=\frac{1}{e^{\omega_k/T}-1}
\end{equation}
is the paramagnon occupation number, and the four-dimensional
$\delta$-function describes energy and momentum conservation,
$\delta^{(4)}(q+k_1+k_2+k_3)=
\delta(\omega_q+\omega_{1}+\omega_{2}+\omega_{3})
\delta^{(3)}({\bm q}+{\bm k}_1+{\bm k}_2+{\bm k}_3)$.
The combinatorial factor $\mathcal{S}$ is
due to summation over
paramagnon polarizations. For details of calculation
of the combinatorial factors  see \textit{e.g.} \cite{Peskin}.
For general O(N) group the factor is
\begin{equation}
\label{comb}
\mathcal{S}=2(N+2).
\end{equation}

Application of Fermi Golden rule to diagrams in Fig.~\ref{F3}
gives the following inverse width
\begin{eqnarray}
\label{gi}
\Gamma^{(i)}_{q}(\omega)&=&\frac{16(2\pi)^6\mathcal{S}\beta_0^2}{2\omega}
\int\frac{d^3k_1}{2\omega_1(2\pi)^3}\frac{d^3k_2}{2\omega_2(2\pi)^3}
\frac{d^3k_3}{2\omega_3(2\pi)^3}\nonumber\\
&&\hspace{-15pt}\times[n_{1}n_{2}n_{3}\ \delta^{(4)}(q-k_1-k_2-k_3)\\
&&\hspace{-15pt}+3(1+n_{1})n_{2}n_{3}\ \delta^{(4)}(q+k_1-k_2-k_3)\nonumber\\
&&\hspace{-15pt}+3((1+n_{1})(1+n_{2})n_{3}\ \delta^{(4)}(q+k_1+k_2-k_3)\nonumber\\
&&\hspace{-15pt}+(1+n_{1})(1+n_{2})(1+n_{3})
\ \delta^{(4)}(q+k_1+k_2+k_3)]\nonumber
\end{eqnarray}
Of course Eq.(\ref{gd}),(\ref{gi}) satisfy the relation (\ref{di1}).
Hence the full width (\ref{di1}) reads
\begin{eqnarray}
\label{gg}
\Gamma_{q}(\omega)&=&(1-e^{-\omega/T})\\
&&\hspace{-15pt}\times\frac{16(2\pi)^6\mathcal{S}\beta_0^2}{2\omega}
\int\frac{d^3k_1}{2\omega_1(2\pi)^3}\frac{d^3k_2}{2\omega_2(2\pi)^3}
\frac{d^3k_3}{2\omega_3(2\pi)^3}\nonumber\\
&&\hspace{-15pt}\times[(1+n_{1})(1+n_{2})(1+n_{3})\ \delta^{(4)}(q-k_1-k_2-k_3)\nonumber\\
&&\hspace{-15pt}+3n_{1}(1+n_{2})(1+n_{3}) \ \delta^{(4)}(q+k_1-k_2-k_3)\nonumber\\
&&\hspace{-15pt}+3n_{1}n_{2}(1+n_{3})\ \delta^{(4)}(q+k_1+k_2-k_3)\nonumber\\
&&\hspace{-15pt}+n_{1}n_{2}n_{3}\ \delta^{(4)}(q+k_1+k_2+k_3)]\ . \nonumber
\end{eqnarray}

\begin{figure}[h!]
\includegraphics[width=0.42\textwidth,clip]{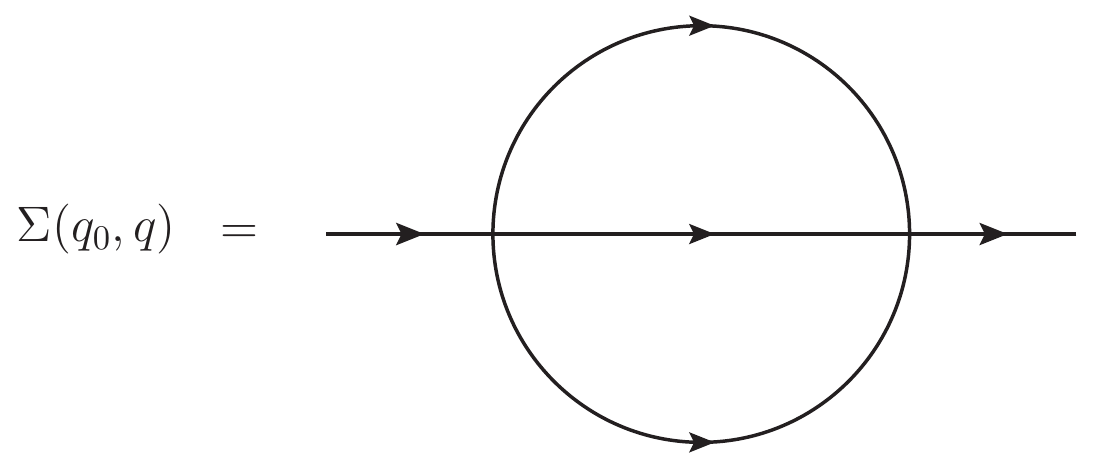}
\caption{ Matsubara self-energy operator.
}
\label{F4}
\end{figure}

One can also derive Eq.(\ref{gg}) more formally starting from the
Matsubara self-energy operator, Fig. \ref{F4},
\begin{align}
\label{ML}
\Sigma(q_0,{\bm q})&=16(2\pi)^2\mathcal{S}\beta_0^2 T^2 \\
\times&\sum_{n_1,n_2,n_2=-\infty}^{\infty}\int\int\int\frac{d^3k_1}{(2\pi)^3}\frac{d^3k_2}{(2\pi)^3}\frac{d^3k_3}{(2\pi)^3}\nonumber\\
\times&\frac{(2\pi)^3\delta({\bm q}-{\bm k}_1-{\bm k}_2-{\bm k}_3)
\delta_{n_0,n_1+n_2+n_3}}
{(k_{01}^2+\omega_{k_1}^2)(k_{02}^2+\omega_{k_2}^2)(k_{03}^2+\omega_{k_3}^2)} \ .
\nonumber
\end{align}
Here $q_0=2\pi T n_0$, $k_{01}=2\pi T n_1$, $k_{02}=2\pi T n_2$, 
$k_{03}=2\pi T n_3$ are Matsubara frequencies, $n_0,n_1,n_2,n_3$ are integer
numbers. Frequencies $\omega_{k_i}$ are given by Eq.(\ref{oq}),
$\delta({\bm p})$ is the $\delta$-function while $\delta_{n,m}$ is the Kronecker
symbol.
Analytic continuation of (\ref{ML}) from $q_0$ to real frequency
together with Eq.(\ref{br}) leads to Eq.(\ref{gg}).
For full details of the analytic continuation see Refs.~\cite{Hou,Pisarski}.

\section{Analysis of Quantum Disordered and Quantum Critical regimes}
\label{QDQC}
\begin{figure}[h!]
\includegraphics[width=0.2385\textwidth,clip]{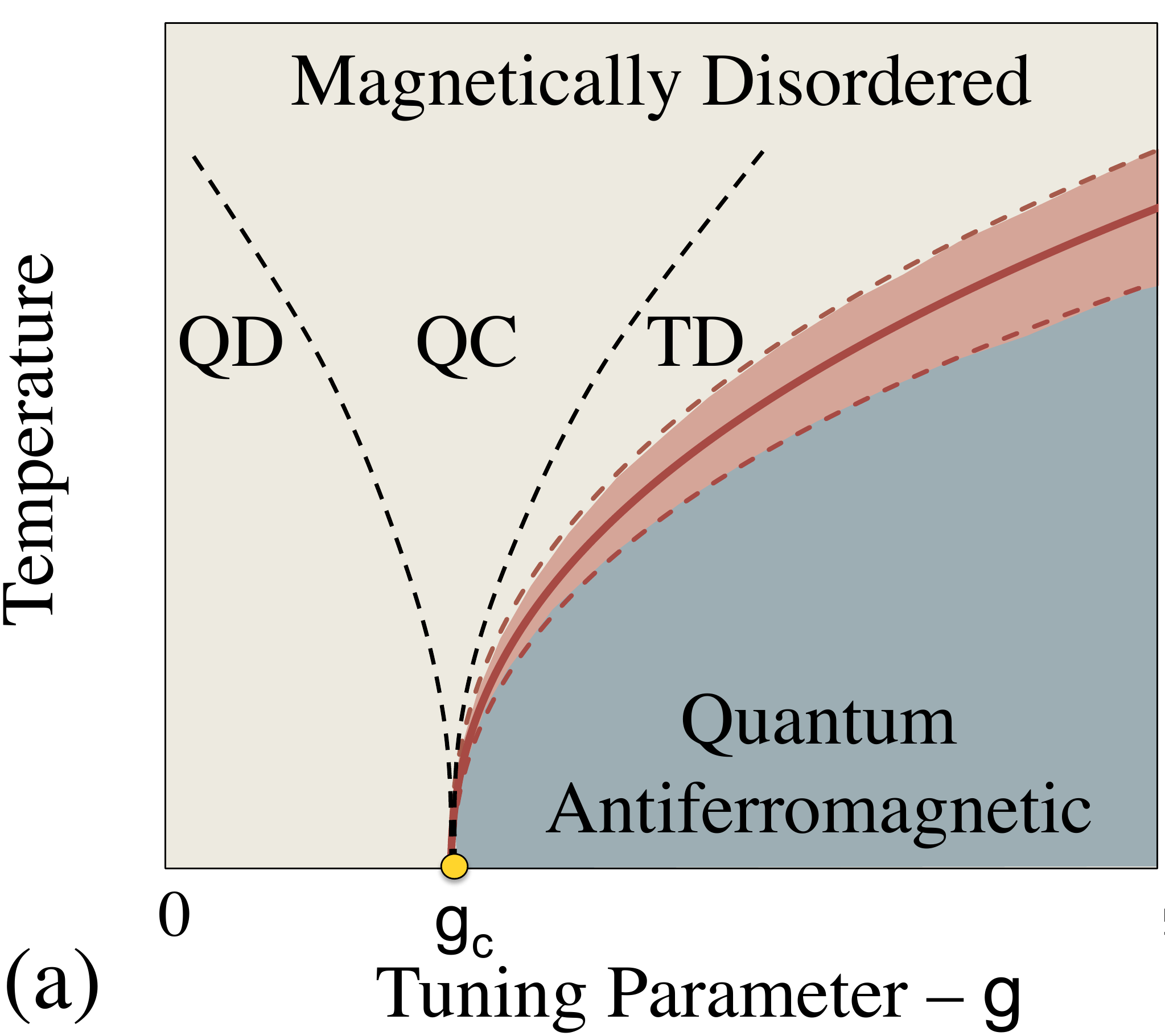}
\includegraphics[width=0.2385\textwidth,clip]{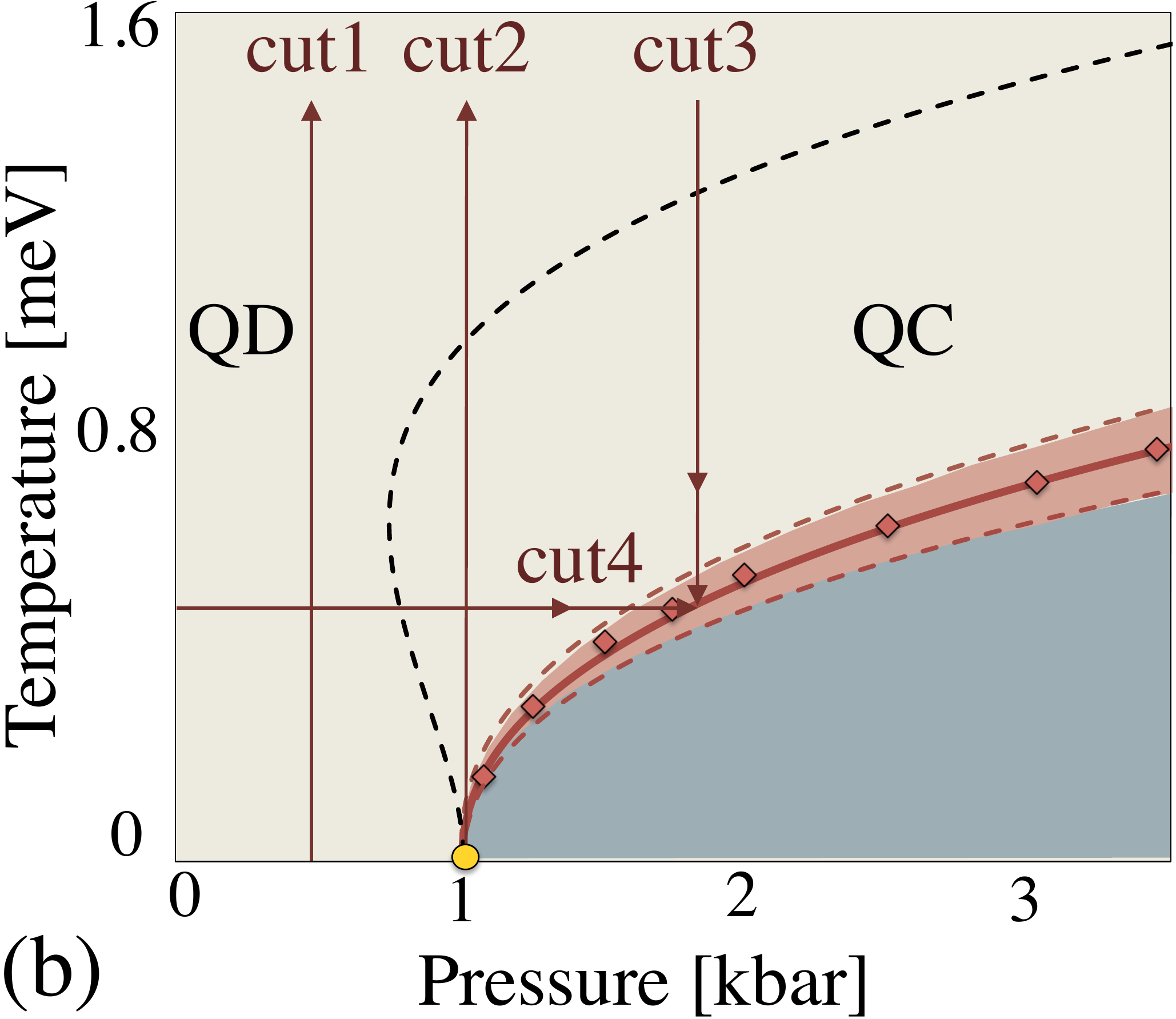}
\caption{Two versions of the phase diagram of a 3D quantum  antiferromagnet.
The N\'eel temperature curve separates
magnetically ordered and magnetically disordered phases.
The light red band around the N\'eel curve indicates the region of dimensional 
crossover.
Panel a: Commonly accepted phase diagram.
The  dashed lines in the magnetically disordered phase indicate smooth 
crossovers between different regimes.
Panel b: The phase diagram derived here, to be specific we use
parameters of TlCuCl$_3$.
The black dashed separates QD and QC regimes. 
The cuts; cut1, cut2, cut3, and cut4 are described in the text.
}
\label{PDC}
\end{figure}

It is well established that critical two-dimensional quantum antiferromagnets
have three different regimes; quantum disordered (QD), quantum critical (QC),
and renormalized classical~\cite{Chakravarty}.
It is widely assumed, see e.g. Ref.~\cite{Sachdev11}, that 
analogously there are three different regimes in the disordered part of the 
phase diagram of a 3D critical antiferromagnet; quantum disordered (QD), 
quantum critical (QC), and thermally disordered (TD).
This is schematically illustrated in Panel a of Fig.\ref{PDC}.
In this section we show that logarithmic corrections 
(running coupling constant) significantly changes this picture.

{ Diagrams contributing to the running coupling constant
and to the self energy are shown in Figs.\ref{Fig22},\ref{Fig33}.
They lead to the following gap equation in the paramagnetic phase~\cite{Scammell}.
}
\begin{eqnarray}
\label{gap}
\Delta^2&=&\gamma^2(p_c-p)\left[\frac{\beta_{\Lambda}}
{\beta_0}\right]^{\frac{N+2}{N+8}}\nonumber\\
&+&8\pi(N+2)\beta_{\Lambda}
\sum_{\bf k}\frac{1}{\Omega_{\bm k}}\frac{1}{e^{\frac{{\Omega}_{\bm k}}{T}}-1} \nonumber\\
\Omega_{\bm k}&=&\sqrt{k^2+\Delta^2+\Gamma^2} \ .
\end{eqnarray}
Here N corresponds to the O(N) group  and 
$\beta_{\Lambda}$ is the running coupling constant
\begin{eqnarray}
\label{RGalpha}
&&\beta_{\Lambda}=\frac{\beta_0}
{1+\frac{(N+8)\beta_0}{\pi}\ln(\Lambda_0/\Lambda)}\\
&&\Lambda=max\{\Delta,T\}\ .\nonumber
\end{eqnarray}
Here $\Lambda_0$ is the ultraviolet normalization point.
 In Eq.(\ref{gap}) we have replaced the general external paramter $g$
to pressure $p$ having in mind further application to TlCuCl$_3$. We will see that in the QD and QC  regimes (away from the Neel curve) 
the width is
always small, $\Gamma \ll \Delta$, therefore $\Omega_{\bm k}$ in
(\ref{gap}) can be replaced by $\omega_{\bm k}$ determined by Eq.(\ref{oq}).

{
In the narrow gap limit, $\Gamma \ll \Delta$, which constitutes  most of
QD and QC regimes, the paramagnon width is determined by the Raman process;
Fig.\ref{F2}b and Fig.\ref{F3}b. Evaluation of  integrals in Eq.(\ref{gg}) 
gives the following explicit answer
\begin{eqnarray}
\label{ngap}
&&\Gamma_{q=0}(\omega=\Delta)=\frac{\pi {\mathcal S}}{2} \beta_{\Lambda}^2T^3\frac{1-e^{-\Delta/T}} {\Delta^2}\mathcal{I}\left(\frac{\Delta}{T}\right)\\
&&\mathcal{I}\left(y\right)=y\frac{6}{\pi^2}\int_{y}^{\infty} dx_1 \int_{y}^{x_1}dx_2 \ n_{x_1}(1+n_{x_2})(1+n_{x_3})\nonumber\\
&&x_3=y+x_1-x_2 \ , \ \ \ n_x=\frac{1}{e^x-1}.\nonumber
\end{eqnarray}
In this equation we substitute the running coupling constant
$\beta_{\Lambda}$ instead of $\beta_0$ in (\ref{gg}), this substitution 
accounts for all RG corrections to  Eq.(\ref{gg}).

It is also useful to calculate the Fermi golden rule $\Gamma_{q=0}(\omega)$ at arbitrary $\omega$.
In this case generally both the Raman  Fig.\ref{F2}b/Fig.\ref{F3}b and the spontaneous
Fig.\ref{F2}a/Fig.\ref{F3}a processes contribute.
Evaluation of  integrals in Eq.(\ref{gg})  gives the following explicit answer.
\begin{eqnarray}
\label{ngap1}
\Gamma_{q=0}(\omega)&=&\frac{\pi {\mathcal S}}{2} \beta_{\Lambda}^2T^3\frac{1-e^{-\omega/T}} {\omega^2}
\left\{\mathcal{I}_b\left(\frac{\omega}{T}\right)+\mathcal{I}_a\left(\frac{\omega}{T}\right)\right\}
\nonumber\\
\mathcal{I}_b\left(y\right)&=&y\frac{6}{\pi^2}\int_{max\{y_0,2y_0-y\}}^{\infty} dx_1 \int_{y_0}^{y-y_0+x_1}dx_2\nonumber\\
&\times& n_{x_1}(1+n_{x_2})(1+n_{x_3})F(x_1,x_2,x_3)\nonumber\\
x_3&=&y+x_1-x_2\ , \ \ \ y_0=\Delta/T \nonumber\\
\mathcal{I}_a\left(y\right)&=&\theta(y-3y_0)y\frac{2}{\pi^2}\int_{y_0}^{\infty} dx_1 \int_{y_0}^{y-y_0-x_1}dx_2
\nonumber\\
&\times& (1+n_{x_1})(1+n_{x_2})(1+n_{x_3})F(x_1,x_2,x_3)\nonumber\\
x_3&=&y-x_1-x_2\ , \ \ \ y_0=\Delta/T \nonumber\\
F(x_1,x_2,x_3)&=&\left\{
\begin{array}{ll}
1 & if \ \ \  x_- \le x_3 \le x_+ \\
0 & otherwise
\end{array}\right.\nonumber\\
x_-&=&\sqrt{\left(\sqrt{x_1^2-y_0^2}-\sqrt{x_2^2-y_0^2}\right)^2+y_0^2}\nonumber\\ 
x_+&=&\sqrt{\left(\sqrt{x_1^2-y_0^2}-\sqrt{x_2^2-y_0^2}\right)^2+y_0^2}
\end{eqnarray}
Of course at $\omega=\Delta$ Eq.(\ref{ngap1}) coinsides with Eq.(\ref{ngap}). 
}
{ It is worth noting that the coupling $\beta_\Lambda$ runs with energy scale $\Lambda=\max\{\sqrt{\omega^2-q^2},T\}$.
}

\subsection{Quantum Disordered Regime}
Consider cut1 in the QD regime of the phase diagram, Panel b of
Fig.\ref{PDC}.
At low temperatures, deep in the QD regime where  $e^{-\Delta/T}\ll 1$, the gap
determined by Eq.(\ref{gap}) is practically equal to its value
at zero temperature. 
Direct evaluation of  the integral in Eq.(\ref{ngap}) gives
\begin{align}
\label{gcut1}
\frac{\Gamma_{q=0}(\omega=\Delta)}{\Delta}&=\frac{ 3\mathcal{S}}{\pi} 
\beta^2_{\Lambda}\frac{T^2}{\Delta^2} e^{-\Delta/T} \ll 1 \ .
\end{align}

\subsection{Quantum Critical Regime}
\begin{figure}[h!]
\includegraphics[width=0.3\textwidth,clip]{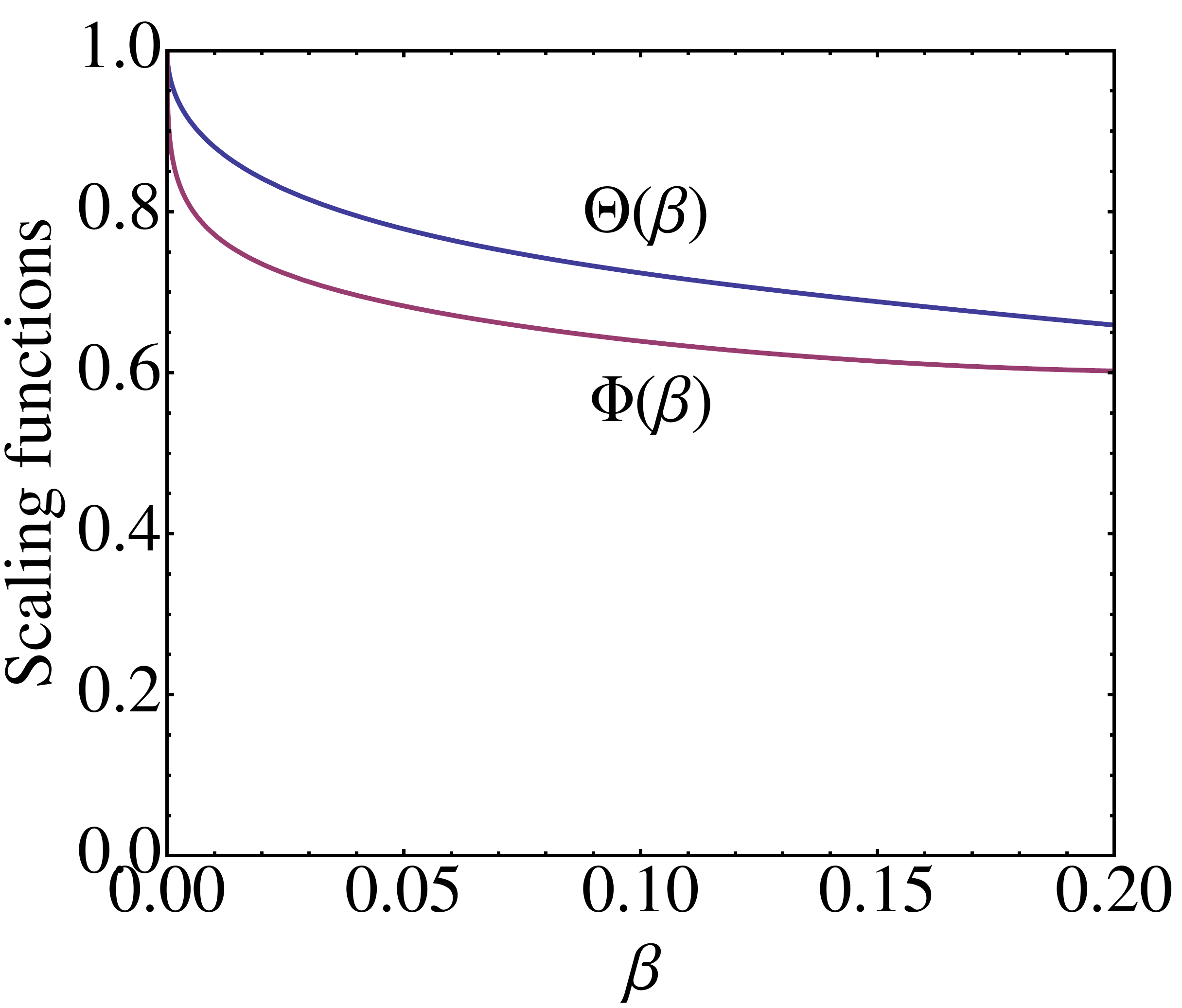}
\caption{Scaling functions $\varTheta(\beta)$ and $\varPhi(\beta)$ 
in Eqs.(\ref{GapScaling}) and (\ref{criticalwidth}) for N=3.}
\label{scalingF1}
\end{figure}

To address the QC regime let us tune to the critical point
by setting $g=g_c$ and increase temperature along cut2
in Panel b of Fig.\ref{PDC}. Solution of Eq.(\ref{gap}) 
in this situation reads
\begin{equation}
\label{GapScaling}
\Delta=T\sqrt{\frac{2(N+2)\pi\beta_{\Lambda}}{3}} \ \varTheta(\beta_{\Lambda}) \ .
\end{equation} 
The scaling function $\varTheta$ is nonanalytic at $\beta \to 0$,
$\varTheta(\beta)= \left(1-\sqrt{\frac{3(N+2)\beta}{2\pi}}+...\right)$, and
therefore deviates from unity noticeably even at small values of the coupling
constant. The plot of $\varTheta(\beta)$ with N=3 is shown in Fig.\ref{scalingF1}.
Hence, using Eqs.(\ref{ngap}) and (\ref{GapScaling}) we find
\begin{align}
\label{criticalwidth}
\frac{\Gamma_{q=0}(\omega=\Delta)}{\Delta}
&=\frac{3{\mathcal S}}{4(N+2)}\beta_{\Lambda} \ \varPhi(\beta_{\Lambda})\ .
\end{align}
Similar to $\varTheta$, the scaling function $\varPhi$, normalized as 
$\varPhi(0)=1$, is nonanalytic in $\beta$. 
The plot of $\varPhi(\beta)$ is presented in Fig.\ref{scalingF1}.
As expected, both $\Delta$, Eq.(\ref{GapScaling}),
and $\Gamma$, Eq.(\ref{criticalwidth}), scale linearly with temperature
along the cut2. However, there is also a logarithmic dependence
related to the coupling constant.
The dependences of  $\Delta$ and $\Gamma$
on the coupling constant are significantly
different. In a very close vicinity of QCP, $T\to 0$, the coupling constant 
(\ref{RGalpha}) is logarithmically approaching zero.
Therefore here $\Gamma \ll \Delta \ll T$.
However, the coupling constant grows with raising temperature  and 
reaches the crossover value $\beta_c$ where $\Gamma \ll \Delta = T$.
The value of $\beta_c$ immediately follows from Eq.(\ref{GapScaling});
for N=3 it is $\beta_c\approx0.23$, and here $\Gamma/\Delta\approx 0.21$.
The crossover value of $\beta$ is sufficiently small, so our approach is
justified. 

\subsection{Crossovers and Contours}
One can define the crossover line between QD and QC regimes by the equality
\begin{equation}
\label{cr1}
\Delta(g,T)=T \ .
\end{equation}
In the QD regime; $\Delta > T$, and in the QC regime; $\Delta < T$.
The crossover line found from Eq.(\ref{gap})
is shown in Fig.\ref{PDC}b by the black
dashed line. It is different from the simple power scaling
indicated in Fig.\ref{PDC}a.
Technically the difference is due to the logarithmic running
of the coupling constant. Physically we say that this difference is due to the system being at its upper critical dimension where there 
are two energy scales; the infrared scale which is equal to temperature and the
ultraviolet one which is determined by position of the Landau pole, see discussion
in Ref~\cite{Scammell}.
As discussed in the previous paragraph, the crossing point
between the black crossover line and cut2 of Fig.\ref{PDC}b
corresponds to $\beta=\beta_c$.

Let us consider now the cut3 in Fig.\ref{PDC}b, which traces from the QD
regime down to the N\'eel phase transition. Along this cut the 
ratio ${\Delta}/{T}$ is monotonically decreasing from: ${\Delta}/{T} \geq 1$ above the QD to QC crossover,
to ${\Delta}/{T}=0$ at the transition. Meanwhile the ratio $\Gamma/\Delta$ is
monotonically increasing. We do not see any fingerprints of a crossover to the ``thermally disordered" regime. From our analysis of the static and dynamic properties we conclude
that separately defining a ``thermally disordered'' regime brings no extra meaning to the phase diagram.
On the other hand, in the very near vicinity of the N\'eel temperature the ratio $\Gamma/\Delta$
becomes equal to unity, and as such brings about a very distinct regime. This regime corresponds to the dimensional crossover to the
``classical critical''  indicated  by the light red band in Fig.\ref{PDC}b.
Our next goal is to describe this crossover.

\section{Decay width expressed in terms of the spectral function.
The Golden Rule of quantum Kinetics.}\label{Conjecture}
Our analysis in previous sections and in particular
derivation of Eq.(\ref{gg}) is based on two grounds: (i) the coupling constant
is small, $\beta \ll 1$, so as to justify the applied perturbation 
theory;
(ii) the paramagnon broadening is small compared to the energy, 
$\Gamma \ll \Delta$,
so that the notion of the thermal occupation number (\ref{BE}) is well
defined.
Close to the N\'eel temperature point (ii) is not valid.
While the coupling constant is still small,
paramagnons become relatively broad as is clearly indicated by 
experiment~\cite{Merchant}.
Note: ``broad'' here means that the width is comparable
or larger than the gap.
Physically, the paramagnons are broad near the N\'eel temperature because 
their gap, Eq.(\ref{oq}), approaches zero as $T \to T_N$. This is the overdamped regime or the ``hot quantum soup''.
In this regime Eq.'s(\ref{BE}) and 
(\ref{gg}) do not make physical sense since a quasiparticle description is 
not well defined. 
Note that quasiparticles with large momentum are still well defined,
$\Gamma_q(\omega=\omega_q) \ll \omega_q$ for sufficiently large q.
The Bose-Einstein occupation number, as presented in 
Eq.(\ref{BE}), explicitly assumes the quasiparticles to be on mass shell; 
$\omega=\omega_{\bm q} = \sqrt{{\bm q}^2+\Delta^2}$. However for broad 
quasiparticles, their dispersion could (crudely speaking) lay anywhere in 
the range $\omega_{\bm q} - \Gamma/2 < \omega < \omega_{\bm q} +\Gamma/2$. 
It is in this sense that the quasiparticle description 
is not valid.
With these considerations in mind, our goal is to develop a theory 
for the regime of large heat bath scattering and subsequent large uncertainty 
in the quasiparticle occupation numbers. We call this the ``hot quantum soup'' 
regime which corresponds to the crossover to the classical critical regime. We do not use the terminology `classical critical' which is appropriate to underline the dimensional crossover; 4D $\to$ 3D, and with it, the unimportance of time. Instead we use the term ``hot quantum soup" to underline the broadening and overdamped dynamics of paramagnons.

To achieve our goal, we first dispense with the Bose-Einstein occupation numbers, and rewrite (\ref{gg}) in terms of
spectral functions.
In the small width regime, point (ii) above, the imaginary part of the retarded Green's 
function follows from Eq.(\ref{gr})
\begin{equation}
\label{gr11}
-\frac{1}{\pi}Im \ G^R(\omega,{\bm q})
=\frac{1}{2\omega_q}\left[\delta(\omega-\omega_q)-\delta(\omega+\omega_q)\right]\ .
\end{equation}
Combining this with (\ref{GS}) we find
\begin{eqnarray}
\label{sca1}
S_{\bm q}(\omega)=\frac{1}{2\omega_q}
\left[(1+n_q)\delta(\omega-\omega_q)+n_q\delta(\omega+\omega_q)\right] \ .
\end{eqnarray}
One can also derive this directly by applying the Fermi golden rule
to the interaction given by the external source (\ref{hint}).
 The first term in brackets in Eq.(\ref{sca1}) describes the creation of a magnon
by the external source, while the second term
in brackets describes a magnon being absorbed from the 
heat bath by the external source.
It is easy to check that using (\ref{sca1}) the width (\ref{gg}) can be 
rewritten as
\begin{align}
\label{GGR1}
\Gamma_q(\omega)&=\mathcal{S}(8\pi)^2\beta^2\frac{(1-e^{-\omega/T})}{2\omega}\int 
S_{{\bm k}_1}(\omega_1)S_{{\bm k}_2}(\omega_2)S_{{\bm k}_3}(\omega_3)\nonumber\\
&\times(2\pi)^4\delta(\omega-\omega_1-\omega_2-\omega_3)
\delta({\bm q}-{\bm k}_1-{\bm k}_2-{\bm k}_3)\nonumber\\
&\times\frac{d\omega_1d^3k_1}{(2\pi)^3}\frac{d\omega_2d^3k_2}{(2\pi)^3}
\frac{d\omega_3d^3k_3}{(2\pi)^3} \ .
\end{align}
An important point is that we can use the general expression (\ref{spectral})
for the structure factor, such that in this form (\ref{GGR1}) does not contain
occupation numbers. The expression is valid for quasiparticles of arbitrary
broadness. In particular, it is valid in the ``hot quantum soup''
regime where quasiparticles are  poorly defined, $\Gamma \gtrsim \omega$.
We call the combinations of these two equations, Eq.(\ref{GGR1}) and
Eq.(\ref{spectral}), the `golden rule of quantum kinetics'.
\begin{figure}[h]
\includegraphics[width=0.42\textwidth,clip]{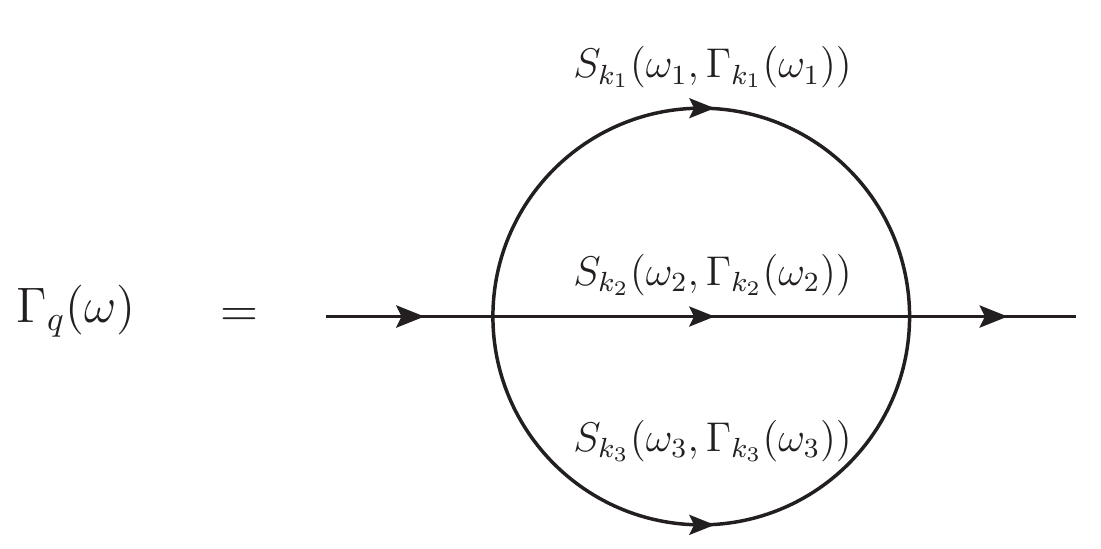}
\caption{ Diagrammatic illustration of
Dyson equation describing
the golden rule of quantum kinetics.
}
\label{F5}
\end{figure}
Self-consistent solution of Eq.(\ref{GGR1}) and
Eq.(\ref{spectral}) is a Dyson-equation-like procedure 
to determine $S_{\bm q}(\omega, \Gamma_{\bm q})$.
Diagrammatically, the Dyson equation is illustrated 
in Fig.\ref{F5}. Most importantly, the solution of the golden rule of 
quantum kinetics gives the structure factor which can be directly compared 
with experiment.
{ Note that Fig.\ref{F5} is not a usual Feynman/Matsubara diagram; the
lines in Fig.\ref{F5} represent structure factors as opposed to Green's functions.
We also comment that Eq.(\ref{sca1}) is used to derive (\ref{GGR1}) in the
narrow line regime, $\Gamma\to0$. However, as soon as one wishes to go beyond simple perturbation theory, and account for the back influence of the decay width on the decay phase space, then Eq.(\ref{sca1}) becomes invalid (generally), and it is Eq.(\ref{GGR1}) and Eq.(\ref{spectral}) that are to be
solved self-consistently.

}

{
Now we can comment on the general structure of our theory and compare with other approaches.
In essence we perform summations of infinite chains of diagrams. The chains of diagrams; those 
for the real part of the self energy and those for the imaginary part of the self energy, are different.
The different chains are dictated by different physics: The real part is dominated
by logarithmic ultraviolet/infrared physics and is related to the logarithmic running 
coupling constant, see discussion after Eq.(\ref{oq}); 
while the imaginary part Eq.(\ref{GGR1}), in the overdamped regime, is dominated by the 
power-divergent, infrared physics. These two different summations cannot be represented as a summation of a single infinite set
of Matsubara diagrams since within the Matsubara technique, the real part and the imaginary part
are treated on an equal footing. And as far as we understand the equation (\ref{GGR1}) cannot be represented within any standard diagrammatic technique.

One of central points of the present work is self-consistent 
Eq.(\ref{GGR1}) for the
spectral function/width. The equation takes care of the infrared, power-divergence in the overdamped regime.
The following points are crucial for the understanding and justification of our approach.\\
{\bf (i)} We assume proximity to the quantum critical point. The proximity implies that
    the logarithmically running coupling constant is sufficiently small  
    to justify truncation of diagrams, $\beta_q \ll 1$.\\
{\bf (ii)} When approaching the Neel temperature the perturbation theory for imaginary part
      breaks down; the width naively calculated using the ``sunset" diagram (analytical continuation 
      of Matsubara) is diverging. This is an infrared power-divergence.  
The failure of the perturbative approach is not a result of the coupling constant becoming large, instead 
      the perturbative approach fails because the gap (=mass) becomes small. The small gap 
      implies the overdamped regime.\\
 {\bf (iii)} Away from the Neel temperature, Eq.(\ref{GGR1}) is equivalent to the simple
   perturbation theory (Fermi golden rule), it gives the same width as
   straightforward analytical  continuation of the ``sunset" Matsubara diagram.\\
 {\bf (iv)} The RG procedure accounts only for the {\it on mass-shell}
contribution to the real part of the ``sunset'' self energy.  However, in our evaluation of the imaginary part of the self energy using Eq.'s(\ref{GGR1}),(\ref{spectral}), we consider both the {\it on} and {\it off mass-shell} contributions. To subsequently find the {\it off mass-shell} contribution
to the real part of the self energy, one can exploit the analytic properties {\it i.e.} Kramers-Kronig relation. This extra step is beyond what is presented in the text, instead the calculation is performed in the Appendix. As expected the {\it off mass-shell} energy dependent contribution is
negligibly small. Furthermore, away from the Neel temperature/overdamped regime, one does not need to consider the {\it off mass-shell} contribution at all. 

 There are approaches to the thermal field theory based on uncontrolled
 truncations of Matsubara diagrams, see e.g. Ref.'s \cite{Hees1,Hees2}. These works do not rely on proximity to a QCP, therefore the coupling constant is, without prior knowledge, large and the truncations uncontrolled. This is not the case in the present work, see point {\bf (i)} above.
Besides that, as already explained, our technique in principle cannot be reduced to a summation
of series of Matsubara diagrams.}

\section{Mathematical analysis of the Golden Rule of quantum Kinetics}\label{GeneralResults}
\begin{figure}[h!]
\includegraphics[width=0.238\textwidth,clip]{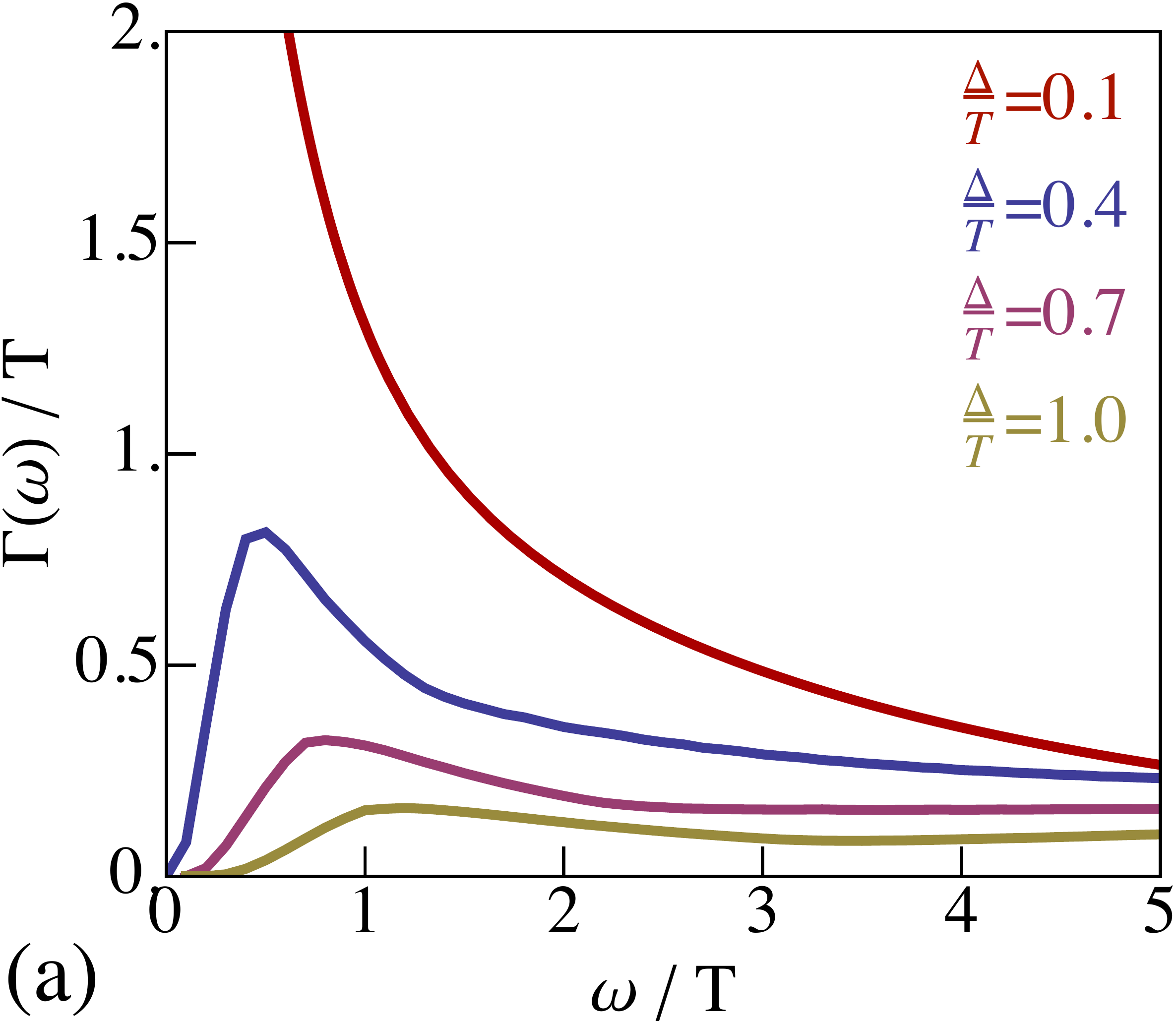}
\includegraphics[width=0.238\textwidth,clip]{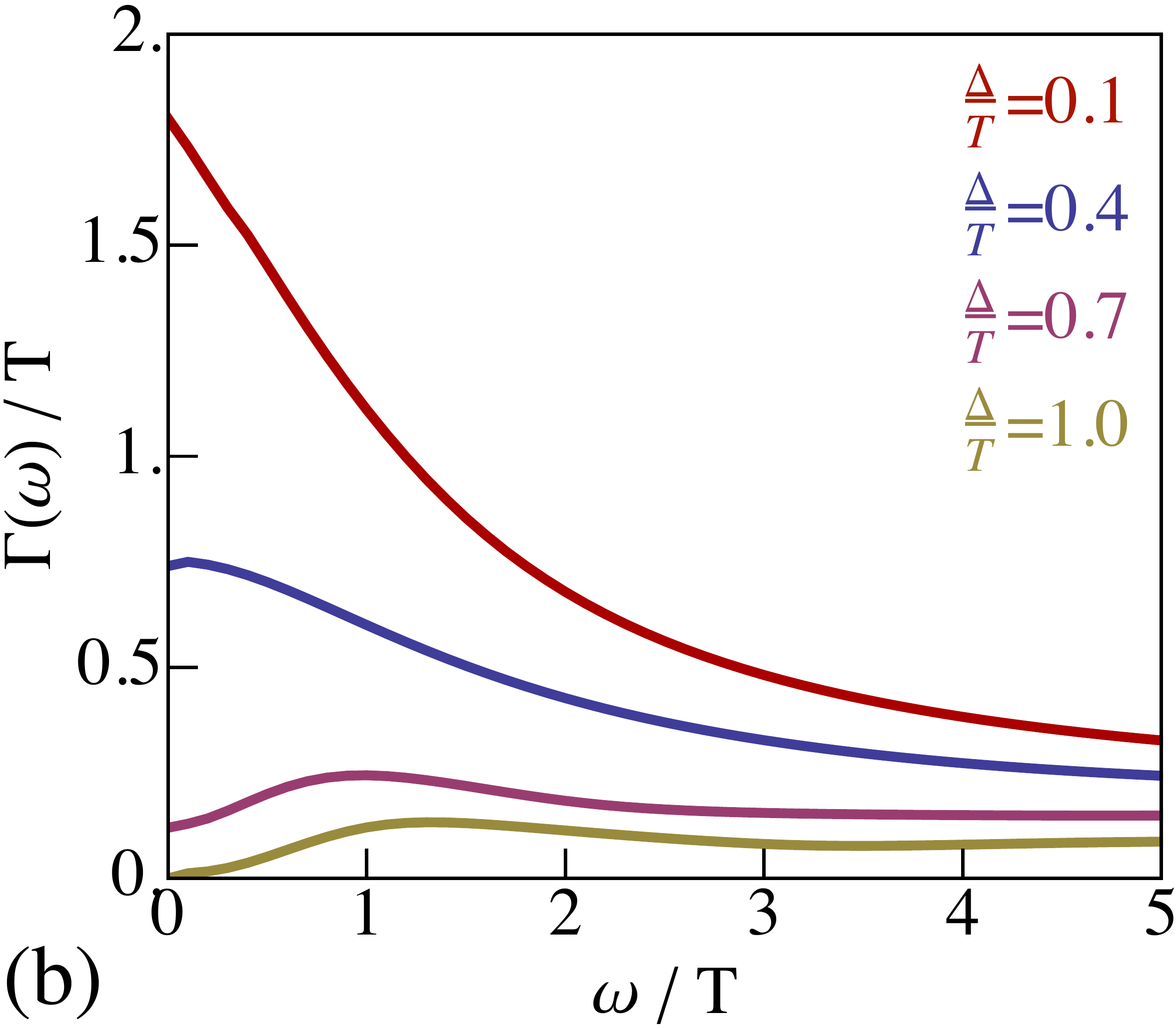}
\caption{Paramagnon width function at zero momentum, $\Gamma_{q=0}(\omega)$,
versus frequency. The function is calculated with the coupling constant 
(\ref{cc}) for different values of the gap $\Delta$.
Panel a: Obtained using the simple Fermi golden rule, Eq. (\ref{gg}).
Panel b: Obtained using the golden rule of quantum kinetics;  
Eqs.(\ref{GGR1}) and (\ref{spectral}).
}

\label{F6}
\end{figure}

{ In this section we provide a general mathematical analysis of the golden rule of quantum kinetics, without reference to any particular system. Our aim is to illustrate the necessity of the non-perturbative resummation of the imaginary part {\it i.e.} the self-consistent solution of of Eq.(\ref{GGR1}) and Eq.(\ref{spectral}). To this end
we disregard the RG running of the coupling constant and set it to
\begin{equation}
\label{cc}
\beta=0.2.
\end{equation} 
In the next section we will again account for the RG running. 

Fortunately the most singular integrations in Eq.(\ref{GGR1}) can be
performed analytically. To avoid long equations here we present the
answer only for $q=0$
\begin{eqnarray}
\label{GGR2}
&&\Gamma_{q=0}(\omega)=\frac{\mathcal{S}\beta^2}{\pi}
\frac{(1-e^{-\omega/T})}{\omega}\int_{-\infty}^{+\infty}d\omega_1d\omega_2
\int_0^{+\infty}dk_1^2dk_2^2\nonumber\\
&&\times\int_{(k_1-k_2)^2}^{(k_1+k_2)^2}dk_3^2 \
 S_{{\bm k}_1}(\omega_1)S_{{\bm k}_2}(\omega_2)S_{{\bm k}_3}(\omega-\omega_1-\omega_2)
\end{eqnarray}
Numerical evaluation of this expression is straightforward.
}  
Consider cut3 in Fig~\ref{PDC}b; we approach the N\'eel temperature
from the QC regime. Along this cut it is convenient to use temperature as the energy scale, and have
$\omega/T$, $\Gamma_q/T$, $\Delta/T$, and $q/T$ as dimensionless variables.
We remind the reader that paramagnon speed is set to unity, $c=1$, and hence 
$q\to cq$ has dimension of energy. 
To illustrate the use of the golden rule of quantum kinetics, and to contrast 
with the usual Fermi golden rule, we present Fig. \ref{F6} which 
shows plots of the paramagnon width function
$\Gamma_{q=0}(\omega)$ versus $\omega$
for values of $\Delta/T$ ranging from $\Delta/T=1$ to $\Delta/T=0.1$.
The width function 
$\Gamma_{q=0}(\omega)$ calculated  using the Fermi golden rule (\ref{gg})
is shown in Fig. \ref{F6}a,
while the width function
calculated  using the golden rule of quantum kinetics, \textit{i.e.} by iterative
solution of Eq.'s (\ref{GGR1}) and (\ref{spectral}), is shown in Fig. \ref{F6}b.
Of course at small $\Gamma/\Delta$, which here corresponds to large $\Delta$, $\Delta/T\gtrsim 1$, the two methods
must reduce to the same result, and they do so, as is evident from Fig.~\ref{F6}.
They also give the same result  at large values of $\omega$.
On the other hand at small values of $\Delta$ and small $\omega$ the results are very different.
This is not surprising since  the Fermi golden rule assumes the
on-mass-shell notion related to Eq.(\ref{BE}), the notion and the Fermi golden rule
fails at sufficintly small values of $\Delta/T$ where the width is very large, $\Gamma/\Delta >1$.
In particular this results in a formal divergence of $\Gamma$ in the limit $\omega, \Delta \to 0$.
On the other hand the golden rule of quantum kinetics does not require the
on-mass-shell notion and therefore does not suffer from the artificial divergence.
 For the remainder of our  analysis we will use only
the golden rule of quantum kinetics.

\begin{figure}[h!]
 \includegraphics[width=0.2385\textwidth,clip]{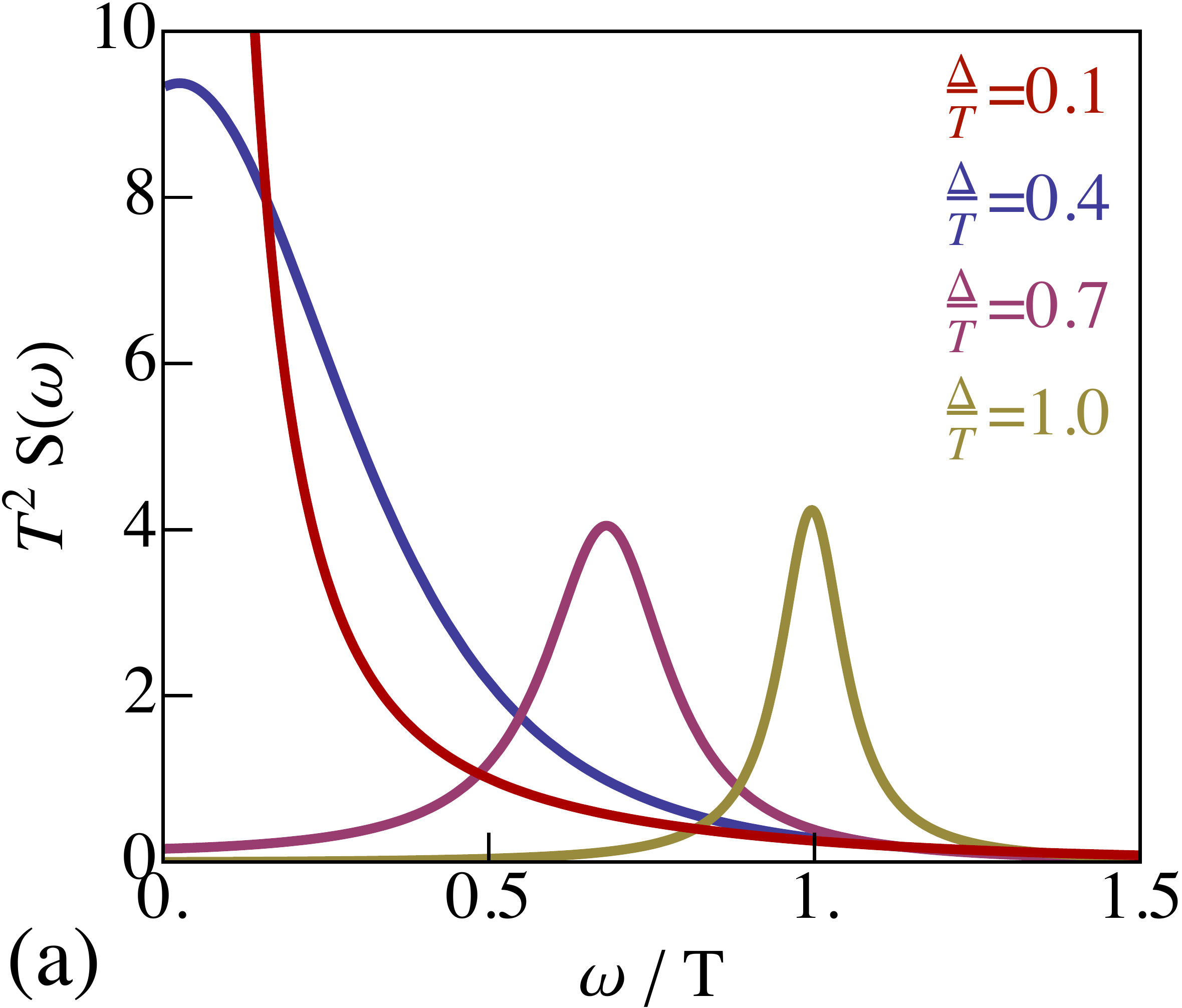}
 \includegraphics[width=0.2385\textwidth,clip]{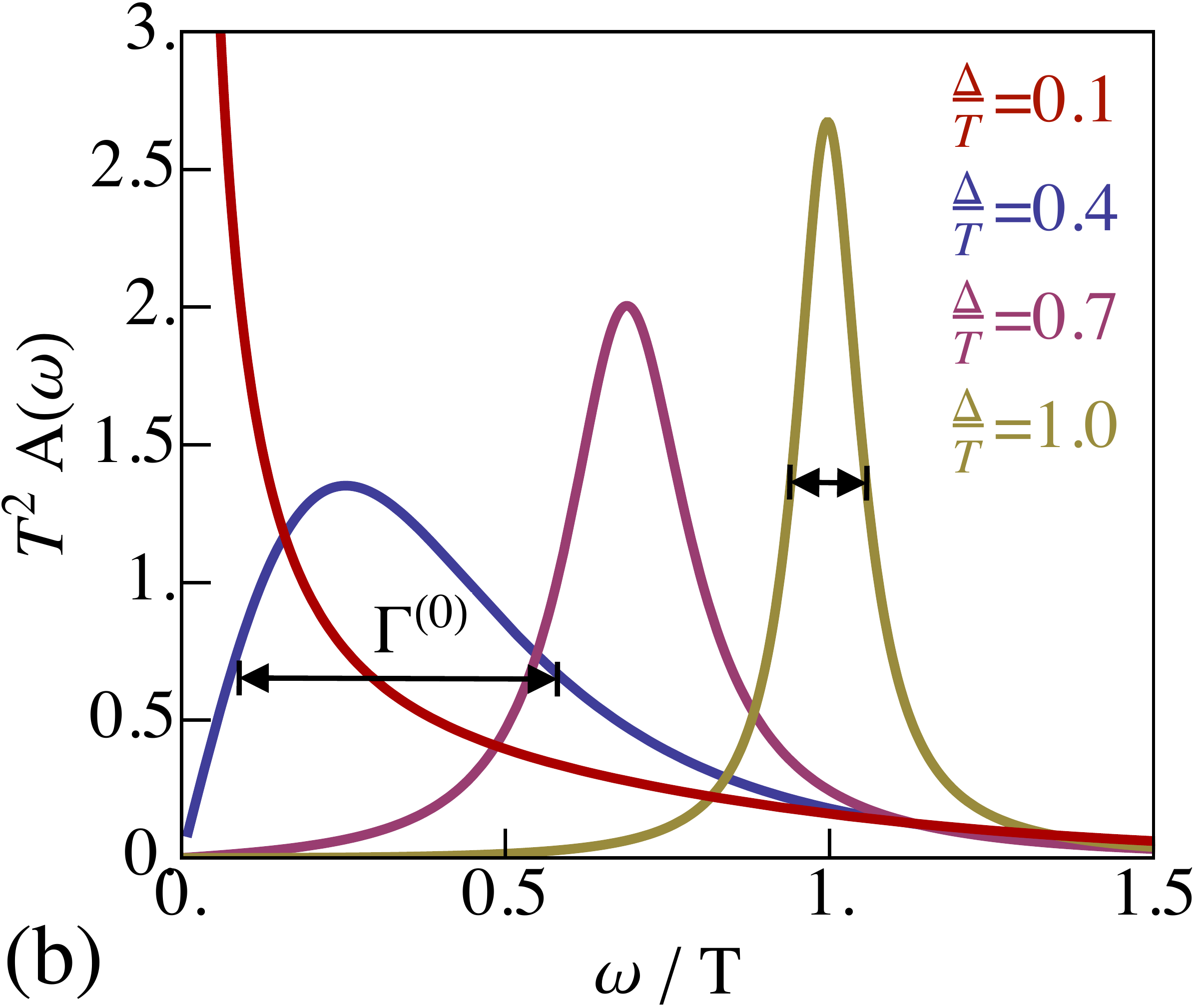}
\caption{ 
Panel a: The structure factor $T^2 S_{q=0}(\omega)$ versus frequency for 
different  values  of the gap $\Delta$.  
Panel b : The spectral density 
$A_{q=0}(\omega)=-\frac{1}{\pi}Im G^R(\omega,q=0)$
versus frequency for different values  of the gap $\Delta$.  
Both $S_{q=0}(\omega)$ and $A_{q=0}(\omega)$ correspond to
$\Gamma_{q=0}(\omega)$ (solid lines) in Figure \ref{F6}.
}
\label{F8}
\end{figure}

The structure factor $S_{\bm q}(\omega)$, as given by Eq.(\ref{spectral}), 
provides a direct physical link to experiment. 
 In Panel a  of Figure~\ref{F8} we present the structure factors 
$S_{q=0}(\omega)$
which correspond to the widths $\Gamma_{q=0}(\omega)$ as given in Figure 
\ref{F6} by solid lines. The structure factor has dimension $[energy]^{-2}$, therefore
similar to other variables in the QC regime we use the appropriate power
of temperature to balance dimension, $S \to T^2 S$.
To supplement the results shown in Panel a of Fig.~\ref{F8}, in 
Panel b of Fig.~\ref{F8} we present plots of the spectral density, 
$A_{\bm q}(\omega)=-\frac{1}{\pi}Im G^R(\omega,{\bm q})$. 
The spectral density $A(\omega)$ is related to
the structure factor according to  Eq.(\ref{GS}).
The spectral density has been used experimentally to determine effective line widths.
In the present analysis, we define $\Gamma^{(0)}_{q}$ to be
the FWHM of the spectral density, which is indicated by the doubled-headed, 
arrowed lines in Fig.~\ref{F8}b.
We stress that $\Gamma^{(0)}_{q}$  has no $\omega$-dependence, but it depends on 
the gap $\Delta$, momentum ${\bm q}$, and temperature $T$. 

At sufficintly small values of $\Delta/T$ the defenition of $\Gamma^{(0)}$ as 
FWHM of the spectral density practically does not make sense, the $A(\omega)$
becomes hugely asymmetric, see the $\Delta/T=0.1$ curve in Fig.~\ref{F8}b. 
This corresponds to the crossover to the overdamped regime,
or in other words to the crossover from quasi-ballistic dynamics to the fully
diffusive one. The crossover value of $\Delta_c$ depends on the value of the
running coupling constant $\beta_{\Lambda}$. The smaller values of $\beta_{\Lambda}$ 
correspond to the smaller  $\Delta_c/T$.
All the available experimental data for TlCuCl$_3$ are in the regime of reasonably well
defined $\Gamma^{(0)}$.

\section{Comparison with experimental data on ${\mbox {TlCuCl}}_3$}\label{Comparison}
The widths of paramagnons $\Gamma^{(0)}_{q=0}$ in TlCuCl$_3$ have been measured 
via inelastic neutron scattering \cite{Merchant}. The data is obtained for
 various values
of $\Delta$ and $T$, spanning the entire phase diagram Fig.\ref{PD}. 
To compare our theory with the data we need to set N=3 and specify 
parameters $\Lambda_0$ and $\beta_0$ in the running coupling constant (\ref{RGalpha}) as well as $\gamma$ and $p_c$ in (\ref{gap}).
The value of $\Lambda_0$ is arbitrary as soon as it is below the position
of the Landau pole, and the value of $\beta_0$ depends of the particular
system/compound and on the value of $\Lambda_0$.
An analysis of the TlCuCl$_3$ data performed in Ref.\cite{Scammell}
shows that for this compound
\begin{eqnarray}
\label{tl}
&&\beta_0=0.23 \ \ \ \ \text{for} \ \ \ \ \Lambda_0 =1 \text{ meV.}\nonumber\\
&& p_c=1.01\text{kbar} \ \ \ \ \ \ \gamma=0.68\text{meV/kbar}^{1/2} \ .
\end{eqnarray}
Note that the analysis~\cite{Scammell} does not include paramagnon widths.
It based solely on the phase diagram and on the data on
values of the quasiparticle
gaps. 

{ Using parameters (\ref{tl}) and the theory developed in the present work
we can calculate gaps.
Let us first consider the cut1 in Fig.\ref{PDC}b and put it at zero pressure
pressure position, $p=0$. The gap and the width along this cut are plotted
in Fig.\ref{phase}. Squares and circles represent experimental data ~\cite{Ruegg2005}
and theory is shown by lines. The gap is determined by Eq.(\ref{gap}) and the
width by Eq.(\ref{ngap}). (Note that Eq.(\ref{gcut1}) is not sufficient since it is
valid only in the regime $\exp(-\Delta/T) \ll 1$.)
The agreement between experiment and theory
for the gap is not surprising, the experimental gap was used in 
Ref.~\cite{Scammell} to determine the parameters (\ref{tl}). Most important, the agreement
for the width is remarkable.
\begin{figure}[h]
 \includegraphics[width=0.2385\textwidth,clip]{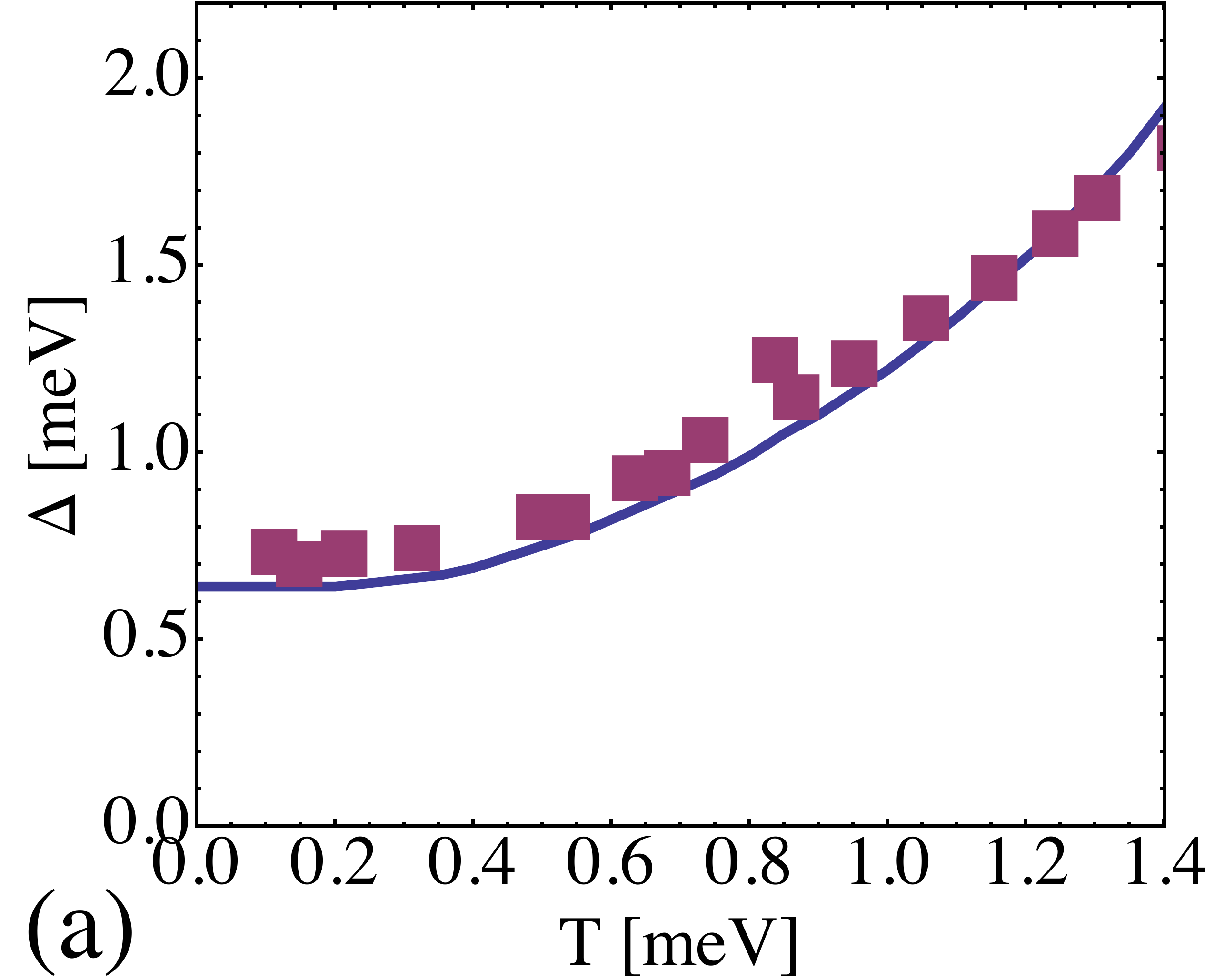}
 \includegraphics[width=0.2385\textwidth,clip]{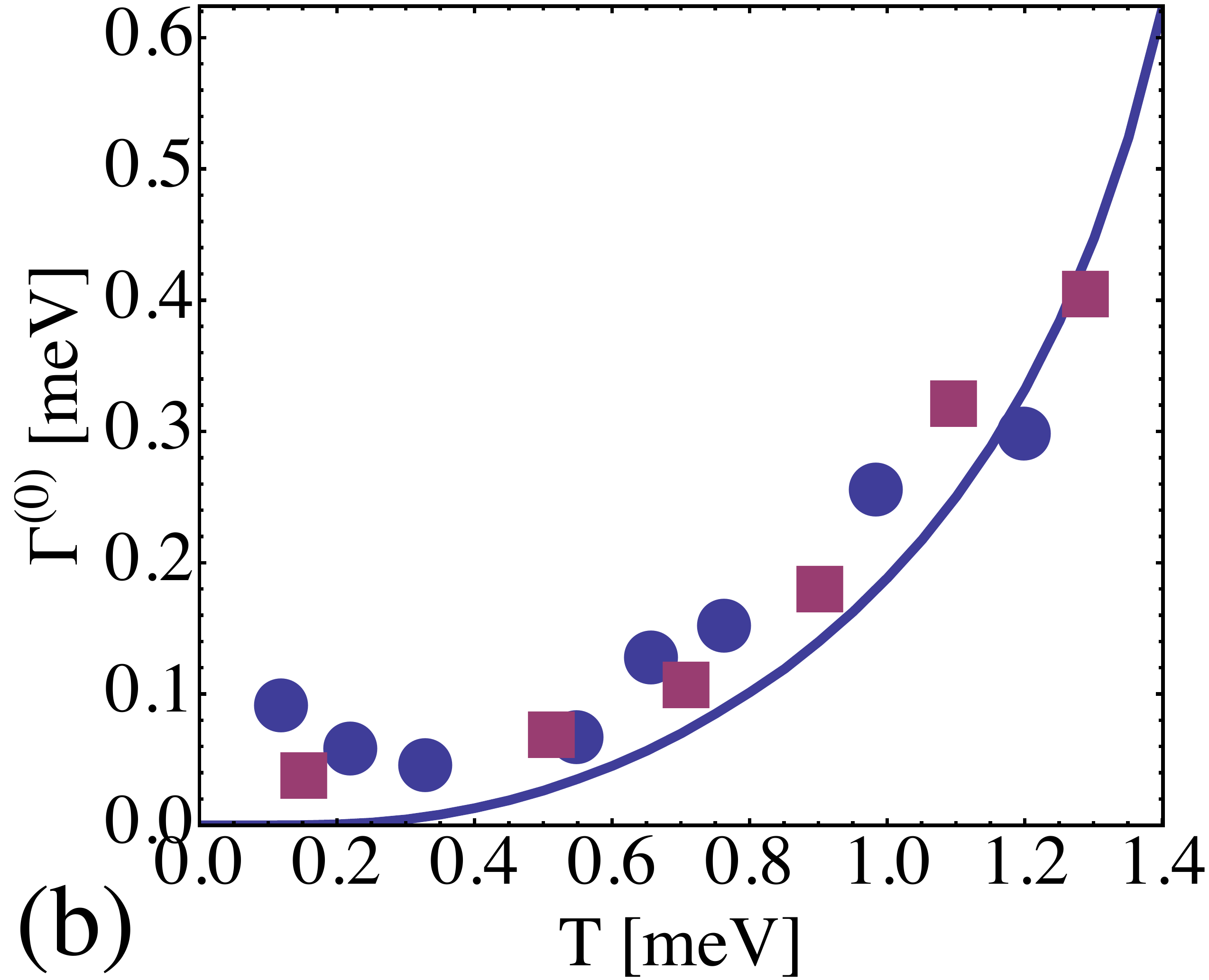}
\caption{The gap (panel a) and the width (panel b) in TlCuCl$_3$ along the
cut1 in the phase digram Fig.\ref{PDC}b. We take the cut at $p=0$ kbar.
Squares and circles represent experimental data ~\cite{Ruegg2005}
and the theory is shown by lines. 
 }
\label{phase}
\end{figure}

Next we consider the cut2 in Fig.\ref{PDC}b, the quantum critical regime.
The gap and the width along this cut are plotted
in Fig.\ref{phase1}. Squares represent experimental data ~\cite{Merchant}
and theory is shown by lines. The gap is determined by Eq.(\ref{GapScaling})
and the width by Eq.(\ref{criticalwidth})
Again, the agreement between experiment and theory is remarkable.
\begin{figure}[h]
 \includegraphics[width=0.2385\textwidth,clip]{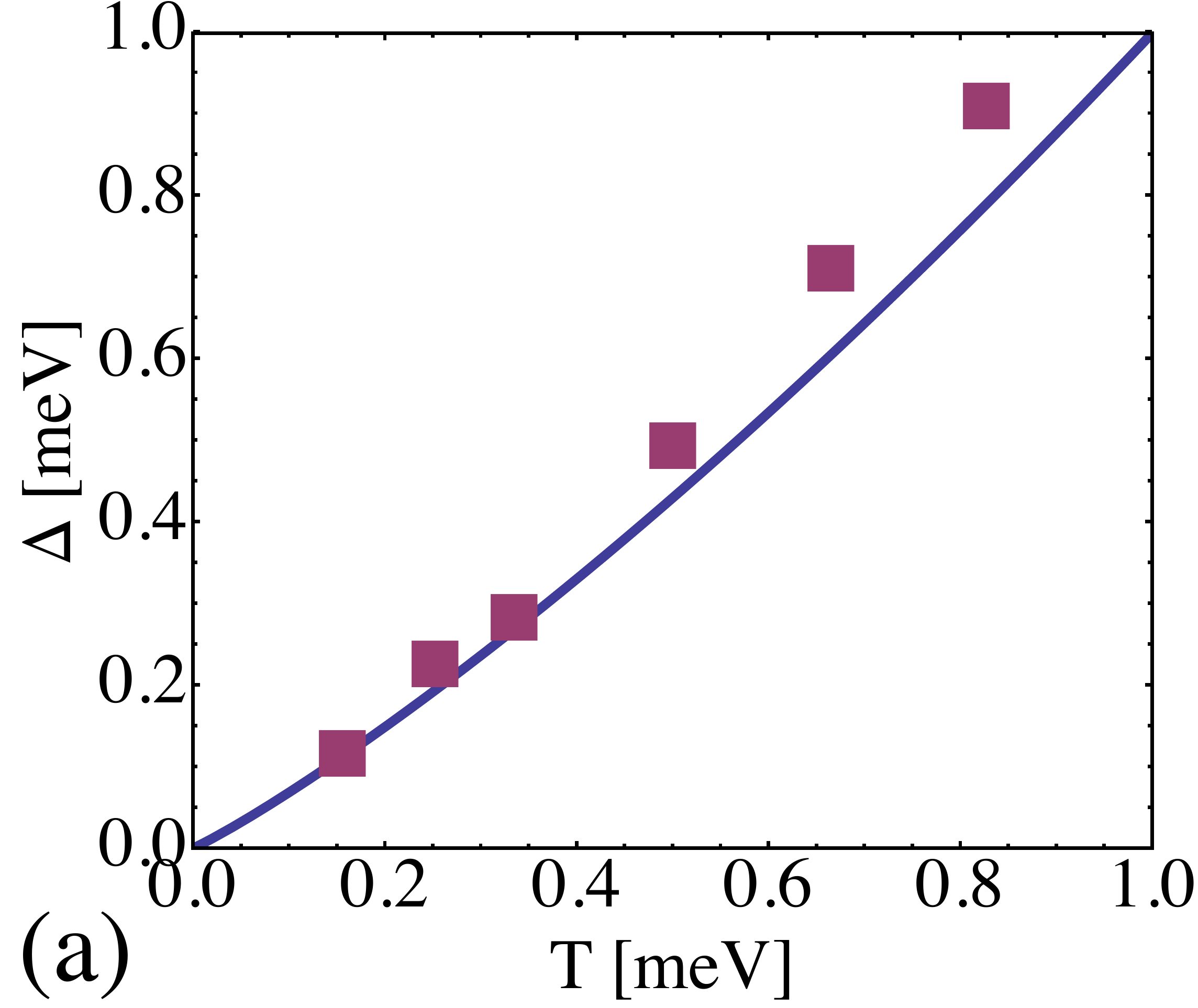}
 \includegraphics[width=0.2385\textwidth,clip]{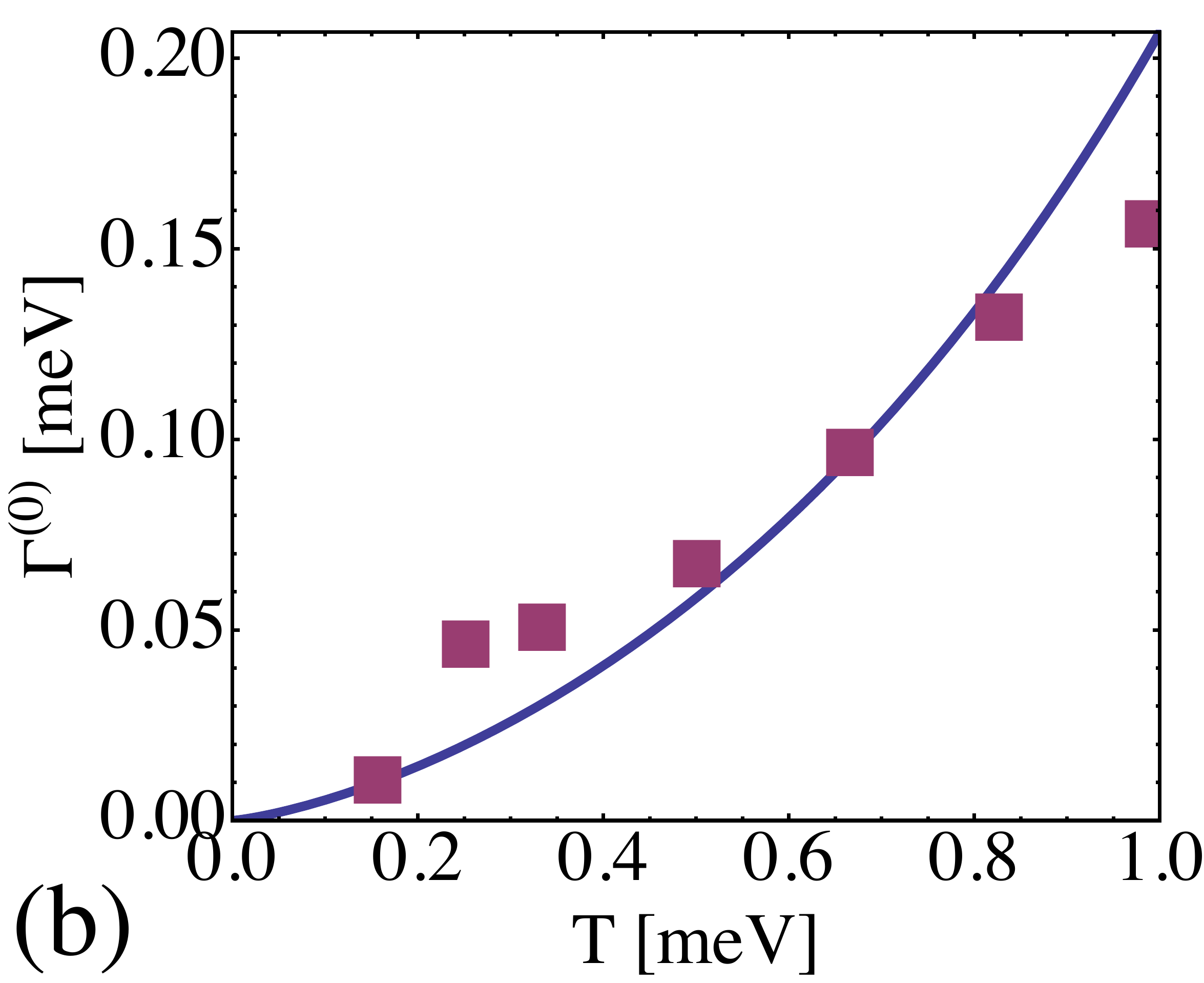}
\caption{The gap (panel a) and the width (panel b) in TlCuCl$_3$ along the
critical cut2, $p=p_c$, in the phase digram Fig.\ref{PDC}b.
Squares represent experimental data ~\cite{Merchant}
and the theory is shown by lines. 
 }
\label{phase1}
\end{figure}

Now we consider cut3 and cut4 in Fig.\ref{PDC}b.
This cuts approach the Neel temperature and hence the 
``simple'' RG used for cut1 and cut2 is not sufficient.
We need RG plus the golden rule of quantum kinetics,
Eqs.(\ref{GGR1}),(\ref{spectral}).
In the vicinity of the Neel temperature spectral lines become asymmetric
and hence the definition of width becomes ambiguous. We use values of
$\Gamma^{(0)}$ defined in section \ref{GeneralResults}.} { In evaluating Eq.(\ref{GGR1}), the coupling $\beta_\Lambda$ formally runs with energy scale $\Lambda=\max\{\sqrt{\omega^2-q^2},T\}$, yet we use $\Lambda=\max\{\Delta,T\}$, which makes a negligible difference \cite{offshell}.}

\begin{figure}[ht]
\includegraphics[width=0.238\textwidth,clip]{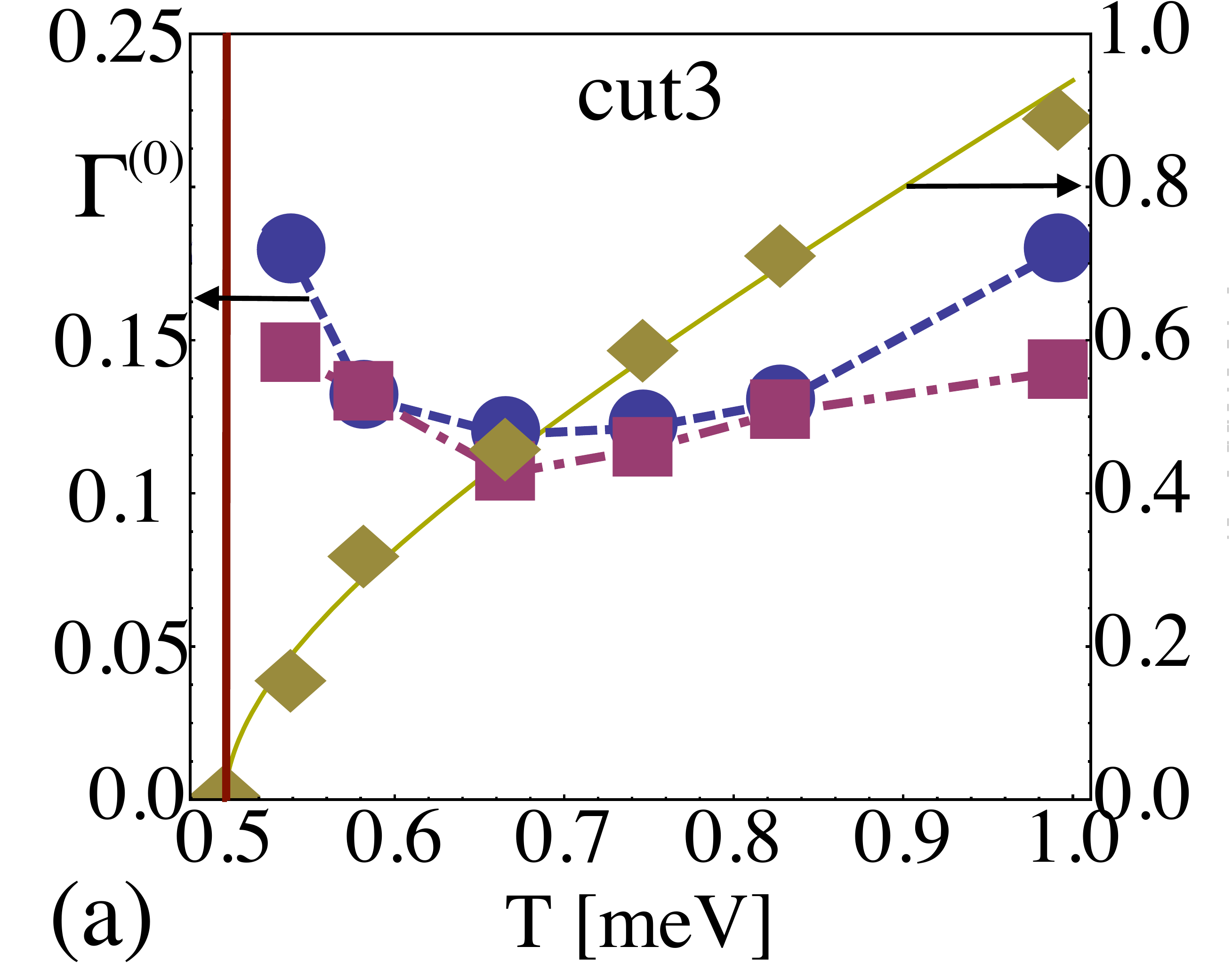}
\hspace*{-0.2cm}\includegraphics[width=0.238\textwidth,clip]{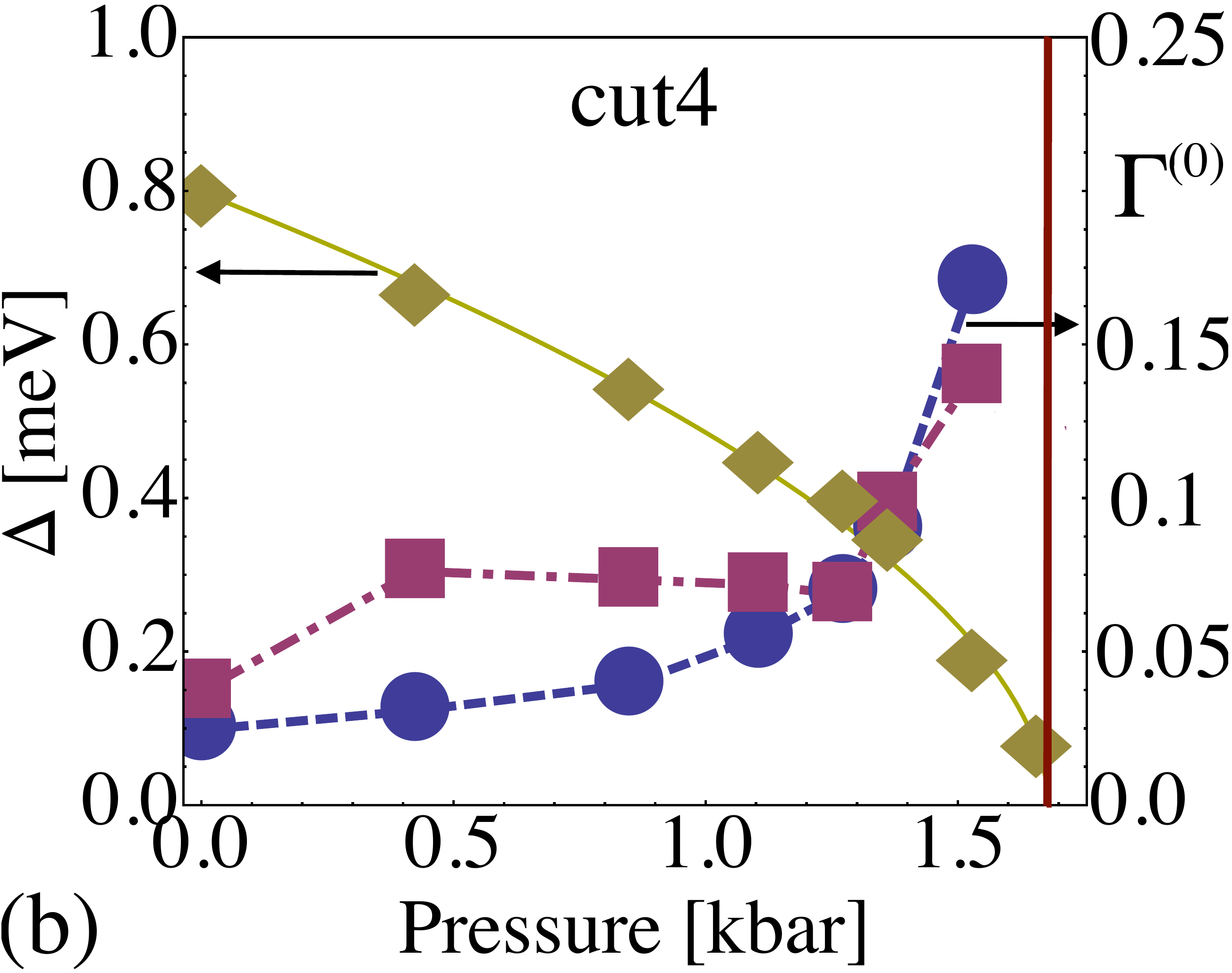}
\caption{
Theoretical and experimental values of the width $\Gamma^{(0)}_{q=0}$ and 
the gap $\Delta$.
Panel a corresponds to the vertical cut3 in  Fig.~\ref{PDC}b, temperature
varies at fixed pressure, p=1.75kbar.
Panel b corresponds to the horizontal cut4 in  Fig.~\ref{PDC}b,
pressure varies at fixed temperature, T=0.5meV.
In both panels blue circles show
theoretical results of the present work while magenta squares
show experimental results of Ref.~\cite{Merchant}. Yellow diamonds show experimental results for the gap \cite{Merchant}. 
Dashed blue and magenta as well as solid yellow lines connecting the points are given just for guidance.
 }
\label{F11}
\end{figure}

In Fig.~\ref{F11} we present theoretical and experimental
values of the width $\Gamma^{(0)}_{q=0}$ and the gap $\Delta$.
Panel a corresponds to the vertical cut3 in  Fig.~\ref{PDC}b; temperature
varies at fixed pressure, p=1.75kbar.
Panel b corresponds to the horizontal cut4 in  Fig.~\ref{PDC}b;
pressure varies at fixed temperature, T=0.5meV.
Agreement between theoretical and experimental widths presented in
 Fig.~\ref{F11}a is very good. This includes the highly nontrivial, hot quantum soup
regime close to the N\'eel temperature where the width calculated via the 
golden rule of quantum kinetics
is different from that calculated via the simple Fermi golden rule.
On the other hand,  Fig.~\ref{F11}b demonstrates a disagreement 
between theory and experiment about factor 2 in the theoretically ``simple''
interval $0 < p < p_c$.
{ In principle one can refer the disagreement to impurities.
However, it is unlikely since the agreement at endpoints of this interval,
$p=0$ Fig.\ref{phase} and at $p=p_c$ Fig.\ref{phase1}, is excellent. 
The reason for the disagreement remains unclear to us.

Finally, to complete this section, in Fig.\ref{Contours}
we present the phase diagram of TlCuCl$_3$ with lines of constant $\Gamma/\Delta$. 
At large T where the running coupling constant becomes large
the lines have small cusps at the QD/QC crossover line ({\it i.e.} when $\Delta=T$).
Of course the cusps are byproducts of the logarithmic RG where the argument
is $\ln(max\{\Delta,T\})$. The magnitude of the cusp indicates the inaccuracy of the RG approach at a given temperature. 
One can consider the line $\Gamma/\Delta=1$ as crossover from the dilute gas
to the hot quantum soup regime.
\begin{figure}[h!]
\includegraphics[width=0.45\textwidth,clip]{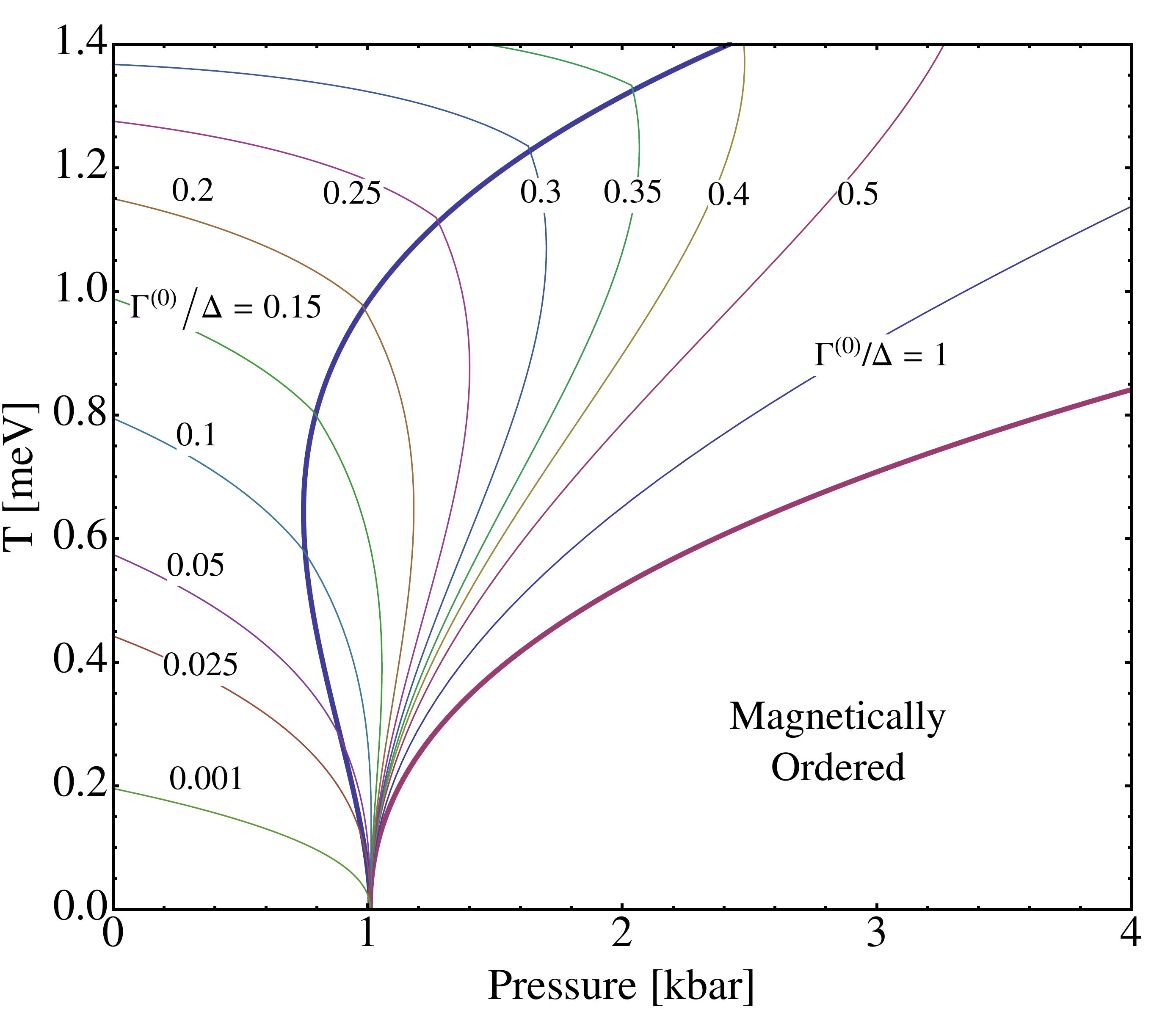}
\caption{Contours of $\Gamma^{(0)}/\Delta$ on the phase diagram of TlCuCl$_3$.
Thick red line is T-Neel, thick blue line is QD/QC
Crossover. All other curves are traces of constant $\Gamma^{(0)}/\Delta$.}
\label{Contours}
\end{figure}
}

There are two comments in conclusion of this section.
(i) Our calculation of widths has no adjustable fitting parameters. 
All parameters were taken from an independent analysis 
Ref.\cite{Scammell} which made no reference to decay widths.
(ii) Calculations performed in this section do not take into account the small 
anisotropy which exists in TlCuCl$_3$. 
It relatively straightforward to account for the anisotropy via introduction 
of an additional anisotropic effective mass as is discussed in 
Refs.~\cite{Kulik, Scammell}. We have performed such a calculation and
checked that the anisotropy does not influence the widths
presented in Figs.\ref{phase},\ref{phase1}, and \ref{F11} beyond a few percent.\\

\section{Conclusions}\label{Conclusion}
We analyze the magnetically disordered phase of 3D quantum antiferromagnets. 
Motivated by observed kinetics of paramagnons in quantum antiferromagnet
TlCuCl$_3$, our analysis is concerned with the non-equilibrium properties: 
paramagnon lifetimes and the neutron scattering structure factor. 
(i) We show that logarithmic running of the coupling constant in
the upper critical dimension changes the commonly accepted picture
of the quantum disordered and quantum critical regimes.
(ii) We calculate paramagnon decay widths in quantum critical and
quantum disordered regimes.
(iii) Close to the Neel temperature the paramagnon width becomes 
comparable to its energy and falls into the hot quantum soup regime
where the quasiparticle lifetimes are very short due to multiple scattering 
from other quasiparticles. 
To describe the ``soup'' we develop a new finite frequency, finite temperature 
technique for a nonlinear quantum field theory; the `golden rule of 
quantum kinetics'. The formulation is generic and applicable to any 
quantum field theory with weak coupling.
(iv) Comparing with data on TlCuCl$_3$ we find an excellent agreement
between theory and experiment.

In the challenging field of many-body quantum systems, a novel technical approach can often help illuminate the physical problem at hand. In this paper we developed a formalism that offers a novel means to calculating non-equilibrium properties of 3+1 dimensional, critical quantum antiferromagnets. Our analysis provides an economical representation, and we hope that the formalism presented here could be applied to other systems of this kind; for example, a wide class of spin dimerised magnetic models.

\section{Acknowledgments}
We thank C. Ruegg, and Y. Kharkov for important comments and discussions. We are especially grateful to B. Normand for critical reading of the manuscript.
The work has been supported by the Australian Research Council, 
grants DP110102123 and DP160103630.

\

\appendix
\renewcommand\thefigure{\thesection.\arabic{figure}} 

\setcounter{figure}{0} 
\section{Non-RG Contribution to the Real Part of the Self-Energy}
\begin{figure*}[t]
\includegraphics[width=0.35\textwidth]{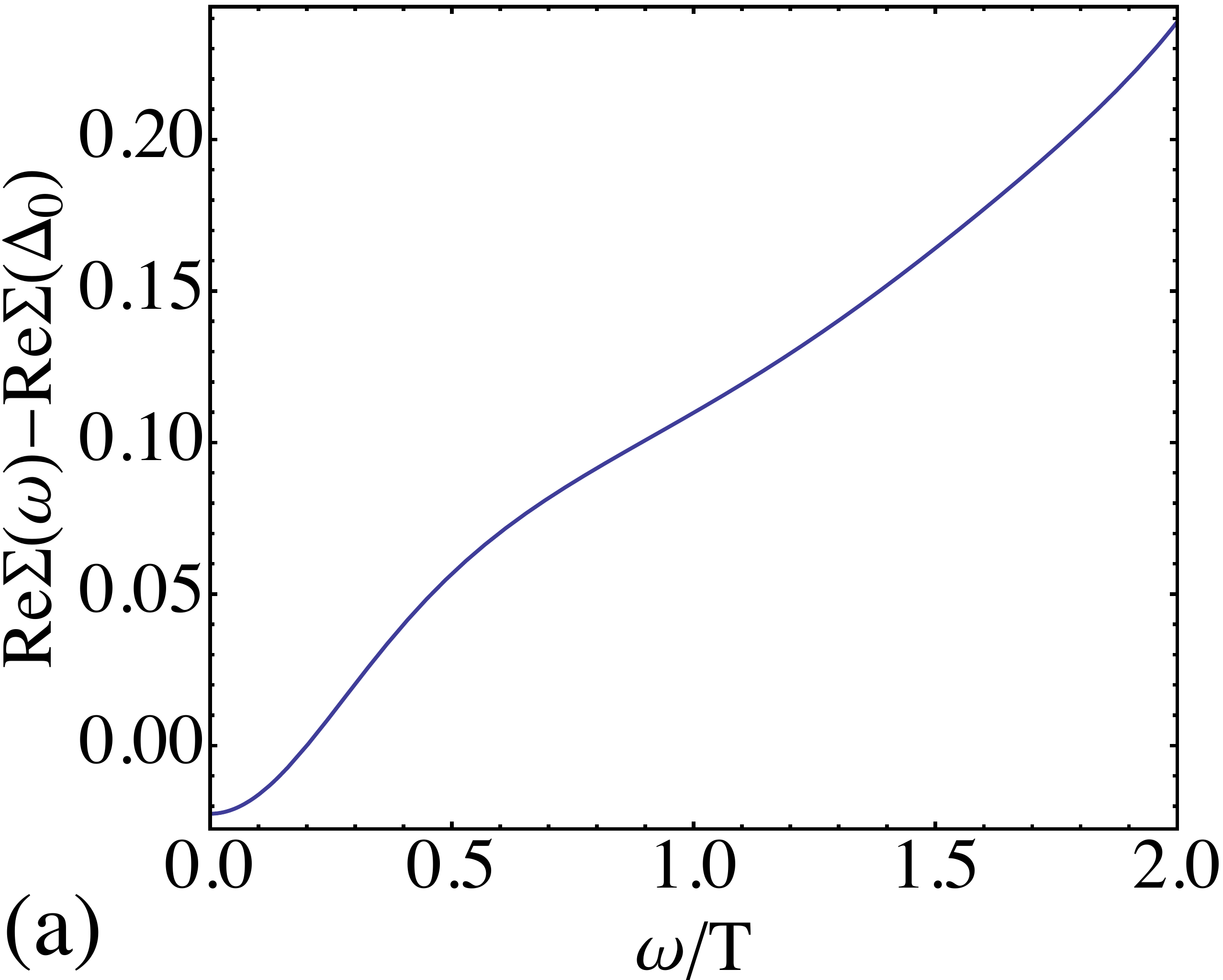}
\hspace{1cm}
\includegraphics[width=0.35\textwidth]{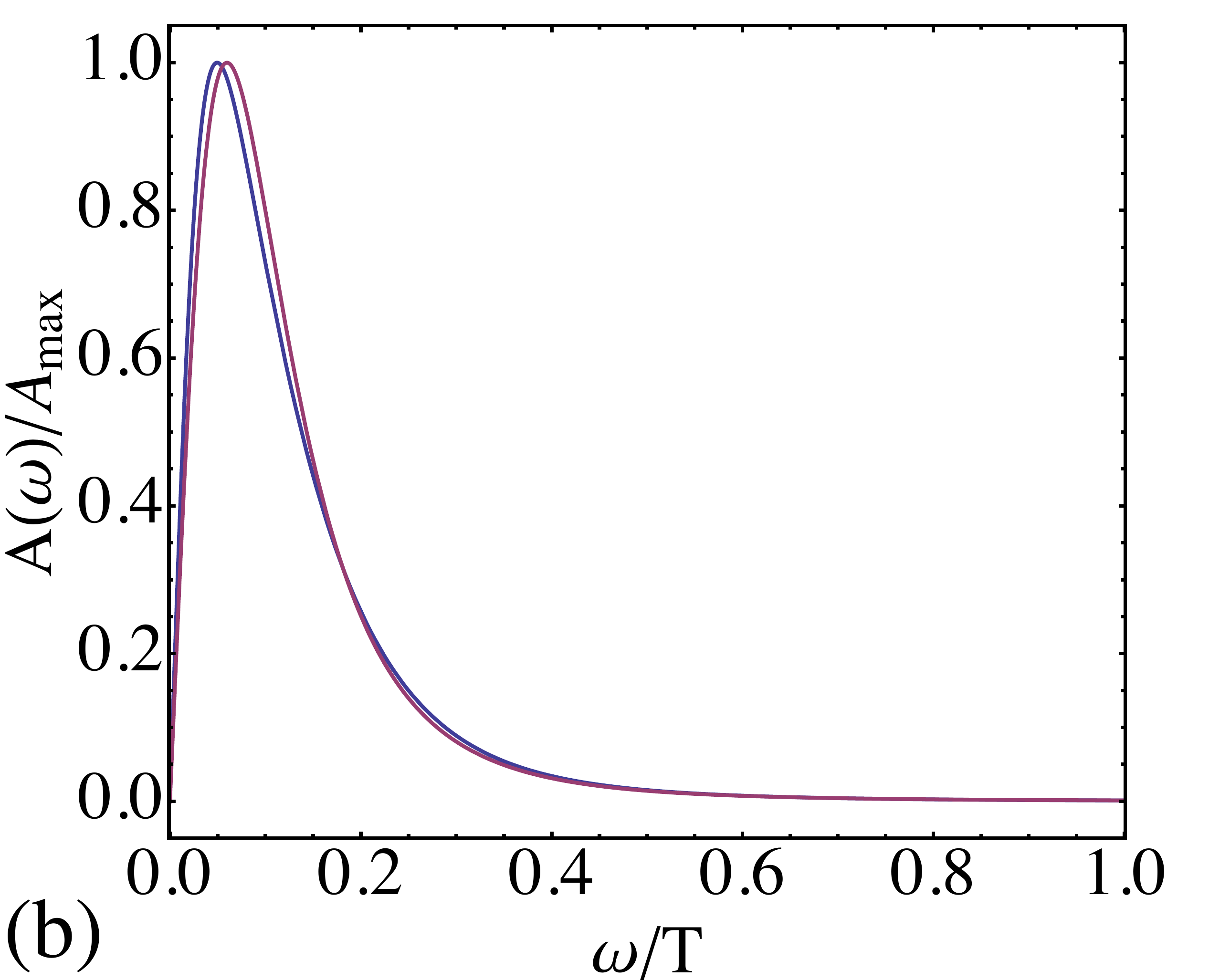}
\caption{Panel a: Frequency dependent correction to mass gap; the non-RG contribution to the real part of self energy. Panel b: The (normalised) spectral density 
$A_{q=0}(\omega)$: (Blue curve) Including the non-RG, frequency dependent correction; (maroon curve) excluding the non-RG, frequency dependent correction.   }
\label{kronig}
\end{figure*}

In the main text we self-consistently solve the golden rule of quantum kinetics Eq.'s (11), (29) to find the imaginary part of the self-energy as well as the structure factor. In doing so, we ignore the small frequency dependence of the real part of the self-energy, $\Re\Sigma_q(\omega)$. Our approximation is equivalent to taking $\Re\Sigma_q(\omega)\approx\Re\Sigma_q(\Delta_0)$, where $\Delta_0$ is the physical mass calculated using RG. In this appendix we take into account the full frequency dependence of the real part of self energy. This is achieved by adding the frequency dependent correction to the mass gap, $\delta\Sigma(\omega)\equiv\Re\Sigma_q(\omega)-\Re\Sigma_q(\Delta_0)$, and solving the following set of equations self-consistently,
\begin{align*}
\tag{A.1}\Delta^2(\omega)&=\Delta_0^2+\delta\Sigma(\omega)\\
\tag{A.2}\Gamma_q(\omega)&=-\frac{\Im\Sigma_q(\omega)}{\omega}\\
\tag{A.3}A_ {\bm q}(\omega)&=
\frac{1}{\pi}\left\{\frac{\omega\Gamma_q(\omega)}{[\omega^2-\left(q^2+\Delta^2(\omega)\right)]^2+\omega^2\Gamma_q^2(\omega)}\right\}.
\end{align*}
Here $\Gamma_q(\omega)$ is defined as in the main text Eq. (\ref{GGR1}), the spectral density $A_{q}(\omega)\equiv(1-e^{-\omega/T}) S_{q}(\omega)$, while the real part is found via analytic properties (Kramers-Kronig relation)
\begin{align}
\notag\Re\Sigma_q(\omega,T)&=\frac{1}{\pi}{\mathcal P}\int^{+\infty}_{-\infty}\frac{\Im\Sigma_q(\omega',T)}{\omega'-\omega}d\omega'\\
\label{Kramers}
\tag{A.4}&=\frac{1}{\pi}{\mathcal P}\int^{+\infty}_{-\infty}\frac{-\omega' \Gamma_q(\omega')}{\omega'-\omega}d\omega'
\end{align}
Here we ignore momentum dependence, which would give some small additional correction. 
Since we already know $\Gamma_q(\omega)$ from solving the golden rule of quantum kinetics, we can use the Kramers-Kronig relation Eq.(\ref{Kramers}) to evaluate the real part.  The results are shown in Fig.\ref{kronig} for the data point $\{\Delta_0,T\}=\{0.2,0.5\}$ meV, with coupling constant $\beta=0.15$. Fig.\ref{kronig}a shows the frequency dependence of the non-RG contribution to the real part of the self energy. Fig.\ref{kronig}b shows the spectral density with and without inclusion of the frequency dependent real part of self energy; blue and maroon curves, respectively. We see that the inclusion of the real part has a negligible influence.


\begin{thebibliography}{99}
\bibitem{ordered phase} 
Note that in  the ordered phase spontaneous decay of a magnon is
generally allowed. For an analysis at zero temperature, we draw the readers 
attention to the work of Ref.'s \cite{Mourigal, Zhitomirsky, Katan}.
\end{thebibliography}
\end{document}